\documentclass[aps,twocolumn]{revtex4-1}
\usepackage{amsmath,amssymb}
\usepackage{slashed}
\usepackage{graphicx}
\usepackage{bm}
\usepackage[top=1.0in,bottom=1.0in,left=1.0in,right=1.0in]{geometry}
\usepackage[colorlinks,linkcolor=blue,citecolor=blue]{hyperref}

\newcommand{\be}{\begin{equation}}
\newcommand{\ee}{\end{equation}}
\newcommand{\bea}{\begin{eqnarray}}
\newcommand{\eea}{\end{eqnarray}}
\newcommand{\ba}{\begin{eqnarray}}
\newcommand{\ea}{\end{eqnarray}}

\begin{document}

\title{Hadronic structure on the light-front II\\ QCD strings, Wilson lines and potentials}

\author{Edward Shuryak}
\email{edward.shuryak@stonybrook.edu}
\affiliation{Center for Nuclear Theory, Department of Physics and Astronomy, Stony Brook University, Stony Brook, New York 11794--3800, USA}

\author{Ismail Zahed}
\email{ismail.zahed@stonybrook.edu}
\affiliation{Center for Nuclear Theory, Department of Physics and Astronomy, Stony Brook University, Stony Brook, New York 11794--3800, USA}

\begin{abstract}
This is the second  paper on hadronic wave functions in the light  front formulation.
We first consider  only confinement effects,  using  the classical Nambu-Gotto string with massive end points in the 
light cone gauge. We derive the light-front Hamiltonian and show how to solve it, using an expansion in suitable basis 
functions.  We next discuss the correlators of Wilson lines on the light front, leading to an effective
Hamiltonian that includes spin effects. At the end, we consider a
separate problem of  instanton-induced interactions, at the origin of the pion as a massless
Goldstone mode on the light front. 
\end{abstract}

\maketitle

\section{Introduction}
 The physics of hadrons is firmly based in  Quantum Chromodynamics,  a theory over half a century old.
 One might think that by now this subject has reached a solid degree of maturity with most issues settled.
 Yet persisting tension remains between the non-perturbative aspects of the theory and empirical measurements
 using inclusive and exclusive processes.
 
More specifically, first principle approaches -- lattice and semi-classics -- are focused on the ground state properties 
of the QCD vacuum,  both using an Euclidean  time formulation. 
 Hadrons are then studied via certain correlation functions. However,  a significant part of the experimental information -- parton distribution functions (PDFs) used in deep inelastic inclusive processes, 
 and distribution amplitudes (DAs) used for exclusive processes --
 are defined  using  light front kinematics, and therefore are not directly accessible by the Euclidean
 formulation.  Only recently, the first attempt to formulate the appropriate kinematical limits~\cite{Ji:2013dva},   and use
 the lattice for calculating the PDFs and PDAs~\cite{Zhang:2017bzy,Alexandrou:2018pbm} were carried out with some success.
 
 Bringing the two sides of hadronic physics together is not just a technical issue related with kinematics.
 Even the main pillars of the theory -- confinement and chiral symmetry breaking -- become contentious. In particular,
 60 years ago Nambu and Jona-Lasinio (NJL) \cite{Nambu:1961tp} have  explained that
 pions are light because they are near-massless vacuum waves due to the spontaneous breaking of chiral symmetry.
 The mechanism creating the vacuum quark condensate and the ensuing organization using chiral perturbation theory,  have since
 been discussed and confirmed in countless papers. More importantly, the QCD vacuum characteristics in the mesoscopic 
 limit reveal  multi-quark correlations captured by  universal spectral fluctuations in the {\em Zero Mode Zone} (ZMZ)~\cite{Verbaarschot:1993pm},
  analogous to the universal conductance fluctuations  around the Fermi surface in dirty metals~\cite{Montambaux:1997svv}, an
  unambiguous  signature of the topological
 nature of the spontaneous breaking of chiral symmetry.
 And yet, parton dynamics is still treated as if the vacuum be ``empty" and 
 quark-partons  massless. There were even suggestions that on the light front
 there are no condensates~\cite{Brodsky:2009zd,Brodsky:2012ku}, 
 although recently these arguments were revisited~\cite{Ji:2020baz}.
 Pions were also argued to be  massless due to other reasons~\cite{DeTeramond:2021jnn},

  Another serious gap  between hadron spectroscopy in the rest frame and on 
  the light front, is due to the complex and dynamical nature of the relativistic boost operator: what
  can be a static potential in one frame, can well become partons on the light front. But even more striking is the difference in the very logical structure of the theory.  In the rest frame spectroscopists 
  start from certain Hamiltonians and 
  derive the wave function, like in atomic or nuclear physics.  On the light front,  phenomenologists mostly
  deal with Parton Distribution Functions
  (PDFs) or Distribution Amplitudes (DAs),  matrix elements of the density matrices or the wave functions
  obtained from experiment or lattice simulations. Few considerations of the light front Hamiltonians and wave functions are
  mostly guessed rather than derived.

  The aim of this work is to $derive$ the light-front Hamiltonian and the corresponding wave functions
  (LFWFs), starting with the most basic meson settings and certain
  nonperturbative dynamics. In this methodical paper we will focus on ``the main components" of the 
  wave functions with zero orbital momentum,  and ignore different  spin structures. Its generalization to full wave functions, with
  all allowed spin and angular momentum values, will be done in the next papers of the series.
  
   One Hamiltonian leads to an  infinite set of wave functions, any of which have infinitely many matrix elements. LFWFs are mutually orthogonal and can be properly normalized. The DAs are normalized
only  to certain empirical constants, like $f_\pi$ for a pion. The PDFs of baryons are traced
over all quarks but one: tracing mixes together all sectors of the wave function, with different
quantum numbers and even the number of partons.  Due to quantum entanglement, it leads to an entropy.

\begin{figure}
		\includegraphics[width=6cm]{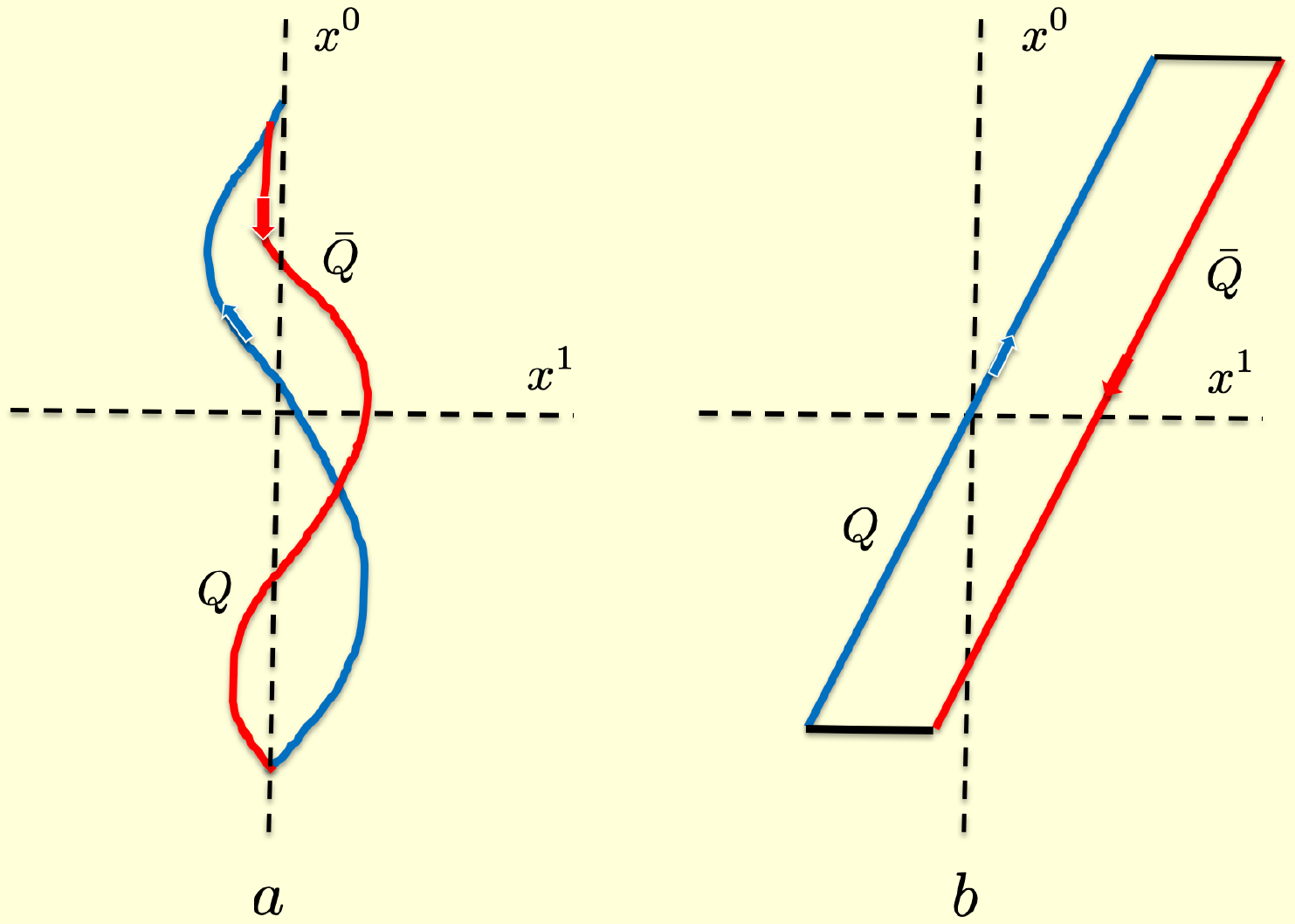}
		\caption{$\bar Q Q$ meson in the rest frame (a) and in the light-front frame (b).}
		\label{fig_qqbar}
\end{figure}

 Deriving these Hamiltonians and solving for  LFWFs is not an easy task, but
going to light front offers certain theoretical advantages. In conventional spectroscopy 
 it is much easier to follow non-relativistic 
 heavy quarkonia which are 
 moving slowly. Central and spin-dependent forces among them can be
formulated in terms of certain universal (flavor-independent) correlators of background fields,
which can be evaluated on the lattice or semiclassically (as we tried to do). 
 Light 
quarks are involved in complicated quantum motion, as depicted 
in the left sketch in Fig.\ref{fig_qqbar}. Usage of nonrelativistic kinetic energy and potentials, as we did   in our paper \cite{Shuryak:2021fsu}, is qualitative at best. 

However, it is improved 
in  the LF  frame, as the motion of all quarks gets ``frozen" (see the right sketch), 
 the distinctions between the heavy and the light quarks basically go away, as both
can be ``eikonalized" and treated in fully relativistic formalism. If so, their interactions can  be deduced  from pertinent Wilson line correlators for any quarks.

\subsection{Light front wave functions}

Light front quantization has a long history with fundamental formulations in QED, atomic and nuclear physics, which we would not review here.
 Instead, we will comment on a recent revival of its use in the
context of hadronic physics. More specifically, we will make use of the  methodical  framework in terms of a
transverse  oscillator basis as  discussed  by Jia and Vary \cite{Vary:2009gt,Jia:2018ary}. Their postulated Hamiltonian 
has led to the light cone wave functions for the pions and rho mesons,  which was shown to be in agreement with a number of experimental results. We will return to  its discussion in section 
\ref{sec_other_H}.

This approach has been extended to  the 3- and 5-quark baryonic sectors by one of us~\cite{Shuryak:2019zhv}, 
addressing the well known puzzle of the isospin asymmetry of the ``antiquark sea". 
Currently this approach has developed to two lines of research. One continues to use
Jia-Vary Hamiltonian with a NJL residual interaction, while the other (called BLFQ collaboration)~\cite{Mondal:2021wfq} trades the 
NJL interaction with an  effective (massive) gluon,  and focus on the two
meson components, $\bar q q$ and $\bar q q g$.

In yet another important theory development, Brodsky and  collaborators \cite{Brodsky:2014yha} have  pointed to  the similarities between 
the QCD  light front quantization for a 2-particle system and the 
``soft wall AdS/QCD" holographic model \cite{Karch:2006pv} for mesons, which elegantly 
reproduces the  Regge dependence on the radial $n$ quantum number of the squared hadronic masses $M^2\sim n$,  via  quantization 
through the  extra holographic coordinate $z$. A special role was
proposed to the ``zeta" combination of the transverse coordinate $ b_\perp$ and the longitudinal momentum fraction $x$, which for mesons is
\begin{equation}
	\zeta\equiv b_\perp \sqrt{x(1-x)}
\end{equation}
It has been suggested  that on the light front it plays 
the same role as the radial  coordinate $r$  in nonrelativistic
quantum mechanics in the rest frame. In other words,   the LFWF wave function
is approximated by a function of this combination only,
$\psi(\zeta)$. 

Furthermore, zeta has been identified with the holographic coordinate $z$ and, on this basis, the corresponding Schrodinger equation was proposed to have the form
\begin{equation} \bigg(-{d^2\over d\zeta^2 }-{1-4L^2 \over 4 \zeta^2}+U(\zeta)\bigg)\psi(\zeta)=M^2 \psi(\zeta)
\end{equation}
with the quadratic  potential 	$U=\kappa^4 \zeta^2+2\kappa^2(J-1)$,  as for the soft wall AdS/QCD.
A similar correspondence was also proposed for baryons, with a ``generalized" $\zeta$ and a corresponding equation,
with a kind of ``supersymmetry"  between mesons and baryons. A physics basis 
for such approximation is approximate {\em constituent quark - scalar diquark symmetry},
suggested in our
work~\cite{Shuryak:2003zi}.

While these works have unquestionably contributed to the rapid and analytic
progress in the field over the past decade, we think it is time to proceed more cautiously,  and 
derive all basic quantities systematically, starting from well established empirical and theoretical  facts. We will
analyze the accuracy of the approximations made, using certain basic examples, focusing 
on the  internal consistency and agreement with the wider set of data, e.g. Regge phenomenology
for principle quantum number $n$ and angular momentum $J$. 

\subsection{The structure of the paper}

  The first problem we address is a basic meson problem with linear confinement.
In section~\ref{confinement} we will derive the light cone Hamiltonian following from the Nambu-Gotto string with massive
ends as a model for a relativistic bound meson by a relativistic string with constitutive quark masses. We will first discuss
the Hamiltonian in 1+1 dimensions and show that its diagonalization leads to the famed $^\prime$t Hooft equation. We will 
then proceed to 1+3 dimensions and derive the corresponding squared mass operator for a relativistic bound meson, which
turns out to be iterative and non-local. In the heavy quark mass limit, we use the semi-classical approximation to detail
the heavy meson spectrum. In general, the iterative and non-local aspect of the squared mass operator can be simplified,
using the einbein trick and minimization, modulo normal ordering.

 The solution of this problem can be done by numerical diagonalization in a suitable basis.
The results obtained for the masses and LFWFs, compare favorably with those
obtained in the rest frame for the spectrum of the same model.  In section~\ref{sec_other_H} we discuss 
other light front Hamiltonians suggested in the  literature.

In our paper \cite{Shuryak:2021fsu} we discussed mesons, in the CM frame. 
In this case the central (and spin-dependent) potentials are derived 
from nonlocal correlators of parallell Wilson lines (and lines with extra fields strengths).
We used certain instanton-based model of the vacuum to evaluate those, and to
relate these potentials and resulting spectra to the mesonic phenomenology. 
In section~\ref{lwilson} we show 
how instantons in the vacuum contribute to the parallell Wilson lines for the non-zero quark modes in  a boosted meson 
on the light front.  The non-zero mode contributions through vacuum tunneling as captured by the spin-flavor dependent
$^\prime$t Hooft interaction, are discussed in section~\ref{t_HOOFT}. In section~\ref{pion} we show how the pion emerges 
on the light cone in the chiral limit with a finite constituent quark mass. The deviation from the chiral limit is in agreement
with the GOR relation for the mass. 
Our conclusions and their discussion are in section~\ref{conclusions}. 

\section{Confinement in the basic relativistic meson problem }

In the first paper of this series
\cite{Shuryak:2021fsu}, we focused on the origin of the central and spin-dependent potentials in CM frame. In particular, we detailed  the lowest states, and used the
nonrelativistic Schroedinger equation not only for heavy quarkonia, but for light mesons as well.  
Now, before we move to the light front quantization, we will discuss a {\it basic} relativisitic 
 problem  of two massive
particles connected by a classical  string, generating a linear confining potential.
We will also discuss not only the low but also high mass excitations.

It is well known from phenomenology,
that the excited mesons (as well as baryons)  form certain Regge trajectories,
relating their masses $M_J$ to angular momentum $J$. For mesons made of   light quarks
they are close to linear 
\be  J=a_M+\alpha' M_J^2
\ee 
where $a_M$ is called the $intercept$ and $\alpha'$ is called the $slope$ of the trajectory,
related to the string tension $\alpha'=1/2\pi \sigma_T$ .
For mesons containing heavy quarks, the trajectories 
are curved. A thorough  discussion of a {\it classical}  model consisting of two rotating masses connected by a string,
is made in~\cite{Sonnenschein:2014jwa}, and provide  accurate Regge-style fits to a number of light and heavy mesons. 

In this  work we rather focus on the {\it radial} excitations of the light mesons, and
consider the relation between the (squared) masses $M_n^2$ and the radial quantum number $n$.
For that, consider the reduced or  {\it half}  of the system in question, with one quark of mass $m_Q$, and {\it half}  the string.
Classically,  the  energy is a sum of kinetic and potential
\be E=V(r)+\sqrt{{\vec p}^2+m_Q^2} \ee
or
\be  (E-V(r))^2=\vec p^2+m_Q^2 \label{eqn_dispersion} \ee 
To quantize it, we use the  standard  substitution of  the squared momentum 
$$ \vec p^2\rightarrow -{\partial^2 \over \partial r^2} -{2\over r} {\partial \over \partial r}+{\hat J^2 \over r^2} $$
with the radial Laplacian (the first two terms) and the angular Laplacian substituted by angular momentum $\hat J^2$. However,
the proper quantization of such a relativistic Klein-Gordon equation is not as simple 
as the nonrelativistic Schreodinger equation. In particular, if one  just takes  the linear form of the potential  $V(r)=\sigma_H r $, 
the problem  is  identical  to that of a constant electric field, in which particle production takes place through the Schwinger mechanism
(Klein paradox), and states with fixed particle number do not stricktly speaking exist. 
We will  return to  the discussion of this equation in Appendix~\ref{KLEIN}, and
here we will use  its semiclassical  treatment by the WKB approximation only.
 
Generically, we have a {\em turning point} $p(r_*)=0$: note that  dynamics is different for $r<r_*$ and $r>r_*$.
 For simplicity, let us start with the massless case $m=0$, in which $E_n-\sigma_T r_*=0$, and $r<r_*$
 where the momentum is real,  and the standard WKB approach can be applied in its original
 form due to Bohr
 \be n \hbar = 4\int_0^{r_*} p(r,E_n) dr \ee 
 (here 4 appears as the integral covers a quarter of the period). 
 
 For $J=0$ and the linear potential $V=\sigma_T r$, this can be analytically calculated, the 
result takes the form 
\bea
n=a_M+ \alpha' E_n^2 \bigg[ \sqrt{1-b^2}+b^2 {\rm log}\bigg({1-\sqrt{1-b^2}\over b}\bigg)\bigg] \nonumber\\
\label{eqn_n_of_E}
\eea
 where the  ``Reggeon slope"  is $\alpha^\prime=1/2\pi\sigma_T$. For two particles
$E_n$ is the total energy of both, and  $b=2m_Q/E_n$.  Here we added a parameter $a_M$, known as a ``quantum shift", which  is not well determined. In~\cite{Sonnenschein:2014jwa} it
 is used  as a free parameter, different for different Regge trajectories.
 We will use the standard WKB recipe taking $a_M=1/2$: as shown  in~\cite{Sonnenschein:2014jwa},
 this value is correct for all light vector mesons $\rho,\omega,K^*,
 \phi$, but take different values for other channels. Of course, confinement is not the only term
 of the Hamiltonian, and shifts are expected. All vector mesons
we will use seem to have such ``residual interactions" to be minimal. They are likely due
to spin forces as we will detail below.
 
\begin{figure}[h!]
	\centering
	\includegraphics[width=7cm]{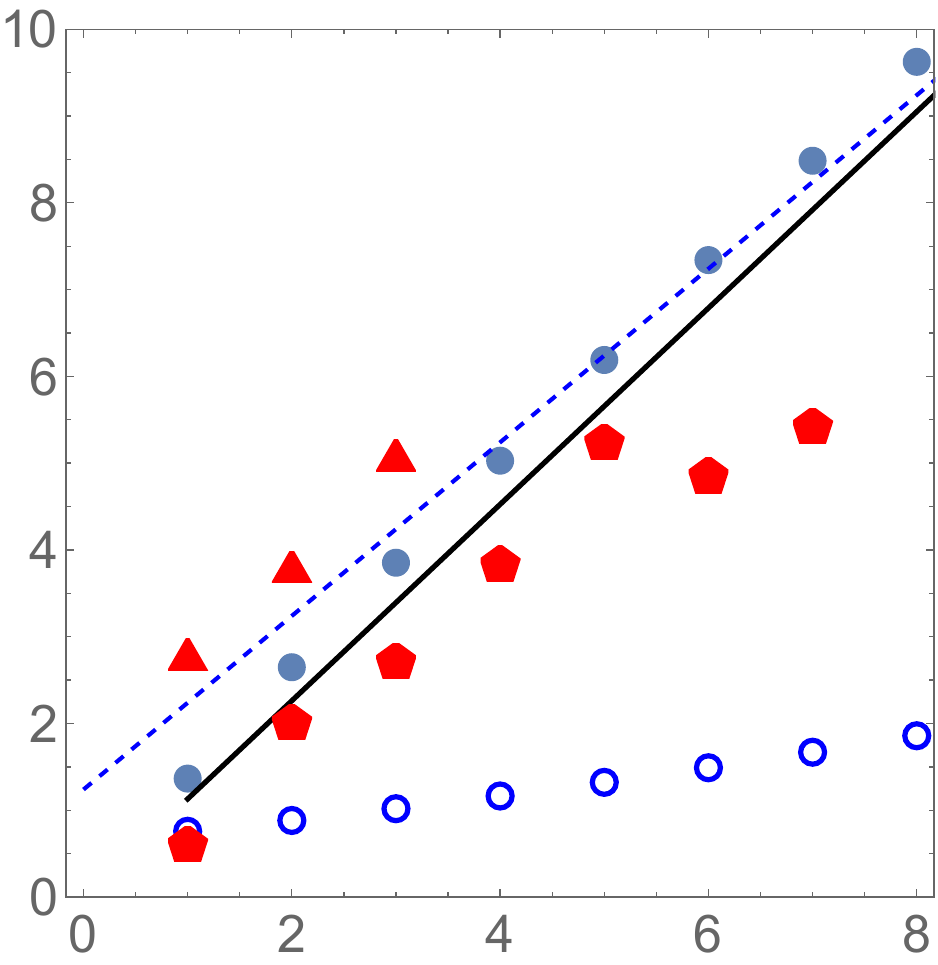}
	\caption{The squared masses of the $n+1$-th states	$E^2_{n+1}\, (GeV^2)$ 
	versus the radial quantum number $n+1$. The closed points are for constituent quarks
	with mass $0.35\, GeV$.
	The solid line corresponds to the massless limit with  $E^2_{n+1}=(n+1/2)/\alpha'$. 
	The red five-polygons are the experimental data for the $\omega$ mesons, and the red triangles are
	for the $\omega_3$ mesons listed in the PDG.
	The
	open points show the corresponding values from the Jia-Vary confining Hamiltonian (\ref{eqn_JV_spectrum}), with
	their recommended value $\kappa=0.227\, GeV$. The dashed line corresponds to the same expression with  $\kappa=0.5\, GeV$.
	}
	\label{fig:basicm2}
\end{figure}

 For arbitrary masses.  the r.h.s. of (\ref{eqn_n_of_E}) is  a complicated function of energy $E_n$,
 which cannot be inverted analytically. Instead of using some approximate formulae, we inverted
 it numerically. In Fig.~\ref{fig:basicm2} the closed points correspond to the squared masses of the radial excitations (in $GeV^2$)
 with $n+1=1-10$ , for a constituent quark mass $m_Q=0.35\, GeV$ and a string tension $\sigma_T=(0.42\, GeV)^2$. The results 
  are  not far from the asymptotic massless formula shown by the thick straight line, so the mass corrections are small. 
  For heavier quarks, the  deviations are larger and  the
 trajectory becomes somewhat curved, as we also suggest below.
 
The red points represent the experimental data, from which we selected two sets of mesons, namely $\omega$
($J=1,l=0$) and $\omega_3$
($J=3,l=2$). Note first, that the numerical value of the string tension, obtained from the quarkonium 
spectra we use, also fits the slope of the Regge trajectories quite accurately. 

Note also, that if one moves the  $\omega_3$ points to the right by two units, they would nearly
coincide with the $\omega$ data points: this means that the 
popular assumption that radial quantum number $n$ and orbital momentum $l$ appear as a simple sum, $n+l$, is approximately correct.
However, in general we do not see why this assumption should be accurate for massive quarks, and plotted them 
without such shift.
Still, it is important that the vertical splitting $M^2(\omega_3)-M^2(\omega)$ is approximately independent of $n$.
In  the  subsequent sections it would be ascribed to two quanta of excitation of the transverse oscillator.

\section{Confinement on the light front }
\label{confinement}

The challenges of the light-front formulation are well known.
A direct boost of the Hamiltonian and wave functions from the rest frame, 
must involve both the Hamitonian and the momentum operators, which
are  hard to achieve. Still, one can compare
certain boost-independent quantities -- excitation masses and
transverse momenta.

\subsection{Confinement in 1+1 dimensions}

As we already mentioned,  chiral symmetry breaking basically solves
the ``mass problem",  giving rise to a  constituent quark mass $m_Q \sim 0.35\,GeV$. 
In the time-like gauge $x^0=\tau$, the free Hamiltonian containing these masses is 

\be
\label{REST}
H\equiv P^0=p_q^0+p_{\bar q}^0=\big(\vec p_q^2+m_Q^2\big)^{\frac 12}+\big(\vec p_{\bar q}^2+m_Q^2\big)^{\frac 12}
\ee
which is the expected result for relativistically moving end points. In the light-cone gauge $x^+=(x^0+x^1)/\sqrt 2=\tau$, 
the on-shell relations reads $2p_{q,{\bar q}}^+p_{q,{\bar q}}^-=m_Q^2$.   The free light-cone Hamiltonian is then
\be
\label{LIGHT}
H\equiv P^-=p^-_q+p_{\bar q}^-=\frac{m_Q^2+\vec p_{q\perp}^2}{2p_q^+}+\frac{m_Q^2+\vec p_{{\bar q}\perp}^2}{2p_{\bar q}^+}
\ee

Confinement is produced by the so called QCD strings (electric flux tubes)  with a string tension 
%
$\sigma_T\approx (420\, MeV)^2\sim 1\, GeV/fm $.
For sufficiently long strings, their  world-sheet dynamics is captured by the classical  and
 universal  Nambu-Gotto action 

\bea
\label{NG}
&&S_P[x]\approx \sigma_T\int_0^T d\tau \int_0^\pi d\sigma \sqrt{(\dot{x}\cdot {x^\prime})^2-\dot{x}^2{x^\prime}^2} \nonumber \\
&&+m_Q\int_0^T d\tau \bigg(\sqrt{\dot{x}^2(\tau, 0)}+\sqrt{\dot{x}^2(\tau, \pi)}\bigg)
\eea
with $x^\mu(\tau, \sigma)$ the string coordinate. The last term represents the  massive end-points, and for simplicity we set the masses equal with $m_Q=m_{\bar Q}$. 
Without the string, the end-points carry momenta $p_{q,{\bar q}}^\alpha=m_Q\dot x_{q,{\bar q}}^\alpha/|\dot x_{q,{\bar q}}|$ with $p_{q,{\bar q}}^2=m_Q^2$. 
The addition of the string will change  the energy, momentum and angular momentum of the $\bar Q Q$ pair,
as we now detail.


As a warm up, 
consider first the simpler case where the string is embedded in  $1+1$-dimensions and ignore the transverse directions.
(\ref{NG}) readily leads to~\cite{Bars:1976nk,Bars:1976re}

\be
\label{LIGHT}
H\equiv P^-=\frac{m_Q^2}{2p_q^+}+\frac{m_Q^2}{2p_{\bar q}^+}+\sigma_T|x_{\bar q}^--x_q^-|
\ee
Using the CM $R= (x_q^-+x_{\bar q}^-)/2$ and relative coordinate $r=(x_{\bar q}^--x_q^-)$,  and their corresponding momenta
$P^+=(p_q^++p_{\bar q}^+)$ and $k^+=(p_{\bar q}^+-p_q^+)/2$, the light cone Hamiltonian (\ref{LIGHT})  yields the squared 
meson mass operator

\bea
\label{LIGHT1}
M^2=2P^+P^-&=&2P^+\bigg(\frac{2m_Q^2P^{+}}{P^{+2}-4k^{+2}}+\sigma_T|r|\bigg) \nonumber \\
&=&\frac {m_Q^2}{\frac 14 -\xi^2}+2\sigma_T|P^+r|
\eea
with the Bjorken $\xi=k^+/P^+=\frac 12+x$  for the fraction of relative momentum carried by the quark at the end-point,
and $|\xi|\leq \frac 12$ or $0\leq x\leq 1$.
Since $k^+$ is canonically conjugate to the relative end-point light-like coordinate $r$ or $[r,k^+]=i$, we can either use the k- or r-representation for the squared mass.
In the k-representation with fixed Bjorken-x, the  coordinate is then the operator $r=id/dk^+$  and the squared mass Hamiltonian (\ref{LIGHT1}) reads
\be
\label{LIGHT2}
M^2=2P^+P^-=
\frac {m_Q^2}{x\bar x}+2\sigma_T|id/dx|
\ee
with $\bar x \equiv 1-x$.

The meson LFWFs and masses follow by diagonalizing (\ref{LIGHT2})

\bea
\label{LIGHT2X}
\bigg(\frac {m_Q^2}{x\bar x}+2\sigma_T|id/dx|\bigg)\varphi_n(x)=M_n^2\varphi_n(x)
\eea
which can be rewritten in  the ${}^\prime$t Hooft equation form~\cite{tHooft:1974pnl}

\bea
\label{THOOFT}
&&M_n^2\varphi_n(x)=\nonumber\\
&&\frac{m_Q^2}{x\bar x}\varphi_n(x)-\frac{2\sigma_T}{\pi}{\rm PV}\int_{0}^{1}dy \frac{\varphi_n(y)-\varphi_n(x)}{(x-y)^2}\nonumber\\
\eea
with the identification of the string with the gauge coupling through  $\sigma_T\equiv g^2N_c/2$ for QCD in 1+1-dimensions, as originally noted in~\cite{Bars:1976nk,Bars:1976re} in the large number of colors $N_c$ limit. The two-dimensional confining potential  in the Bjorken-x representation is

\bea
\langle x||id/dx||y\rangle &=&\int_{-\infty}^{+\infty}\frac {dq}{2\pi}e^{iq(x-y)}|q|\nonumber\\
&\rightarrow& {\rm PV}\frac {-1}{\pi(x-y)^2}+\frac {-1}{\pi x\bar x}
\eea
using the principal value prescription,

\be
{\rm PV}\frac 1{z^2}=\frac 12\bigg[\frac 1{(z+i0)^2}+\frac 1{(z-i0)^2}\bigg]
\ee
The induced self-energy which is {\em negative}  in (\ref{THOOFT})

\be
-\frac{2\sigma_T}{\pi}{\rm PV}\int_{0}^{1}dy \frac{-\varphi_n(x)}{(x-y)^2}
= \frac{-2\sigma_T/\pi}{x\bar x}\varphi_n(x)
\ee
can be made more manifest by recasting the ${}^\prime$t Hooft equation (\ref{THOOFT})  in the equivalent form

\bea
\label{THOOFTX}
&&M_n^2\varphi_n(x) =\nonumber\\
&&\frac{m_Q^2-2\sigma_T/\pi}{x\bar x}\varphi_n(x)-\frac{2\sigma_T}{\pi}{\rm PV}\int_{0}^{1}dy \frac{\varphi_n(y)}{(x-y)^2}\nonumber\\
\eea

 The spectrum following from (\ref{THOOFTX}) admits a  massless 
 mode
 $\varphi_0(x)\rightarrow \theta(x\bar x)$, provided that the {\em current quark mass} $m_Q\rightarrow 0$. (The constituent quark mass in QCD in 1+1 dimensions
 is gauge dependent and divergent).
 The {\em positive} string pair interaction balances  the induced Coulomb self-energy  which is {\em negative}.  
 In QCD in 1+1-dimensions, the massless mode
 appears only in the large number of colors limit owing to the Berezenskii-Kosterlitz-Thouless mechanism.
In massless QCD in 1+3-dimensions, the pion is a true Nambu-Goldstone mode. 

The semi-classical spectrum following from (\ref{THOOFT}) Reggeizes  with a mass gap 

\bea
&&\int_{x_-}^{x_+}dx \bigg(M_n^2-\frac{m_Q^2}{x\bar x}\bigg)=\nonumber\\
&&M_n^2-m_Q^2\,{\rm ln}\bigg(\frac{x_+\bar{x}_-}{x_-\bar{x}_+}\bigg)=2\pi\sigma_T n
\eea
with the turning points
\be
x_\pm=\frac 12\bigg(1\pm \bigg(1-\frac{4m^2_Q}{M_n^2}\bigg)^{\frac 12}\bigg)
\ee
and with  $M_n\geq 2m_Q$. The mass gap vanishes  for $m_Q\rightarrow 0$ with  a radial Regge trajectory $M_n^2= n/\alpha^\prime$,
and $\alpha^\prime=1/2\pi\sigma_T$ the slope of the open bosonic string as it should.  At large $n$, the $^\prime$t Hooft equation (\ref{THOOFT}) can be solved
semi-classically giving the light cone wavefunctions $\varphi_n(x)\approx \sqrt 2 \,{\rm sin}((n+1)\pi x)$~\cite{tHooft:1974pnl}.
In the massive case, the  Regge trajectory is
modified  to $M_n^2\approx n/\alpha^\prime +2m_Q^2\,{\rm ln}\,n$.

\subsection{Light-front Hamiltonian for a string in 1+3 dimensions}

  We now return to our main problem, the mesonic Hamiltonian and wave functions on the light front. We use
 the  string (\ref{NG}) in  the light cone gauge in 1+3-dimensions. Ignoring the string vibrations
(Luscher term and its corrections in higher order) the  result  can be read off 
from  (\ref{LIGHT2}) 

\begin{widetext}
\be
\label{LIGHT3X}
M^2=2P^+P^-=
\frac {m_Q^2+k_\perp^2}{x\bar x}+2\sigma_T\bigg(|id/dx|^2+\frac{P^{+2}x_\perp^2}{\gamma^2}\bigg)^{\frac 12}
\ee
with the Lorentz factor $\gamma=P^+/Mv\rightarrow \infty$ as  $v\rightarrow c$ gets close to  the light cone,

\be
\label{LIGHT3}
M^2=2P^+P^-=
\frac {m_Q^2+k_\perp^2}{x\bar x}+2\sigma_T\bigg(|id/dx|^2+M^2 x_\perp^2\bigg)^{\frac 12}
\ee
\end{widetext}
 Again, the transverse coordinate and momenta $x_\perp, k_\perp$ are conjugate 
 and (for fixed Bjorken-x) it is appropriate to use $\vec x_\perp=i \vec \nabla_\perp$ to diagonalize the squared mass operator in full momentum representation.
 The generalization of (\ref{LIGHT2}) in 1+1 dimensions  to (\ref{LIGHT3}) in 1+3 dimensions, was also noted in~\cite{Pirner:2009zz}.
 
The squared mass operator (Hamiltonian) is now given in terms of a non-linear 
differential operator. It is a symbolic form since one still has to define certain procedures for calculating its matrix elements, which can be done only
{\em modulo ordering ambiguities}. A good test for these procedures is provided by a requirement that the mass spectrum be the same as in the CM frame. In particular, the semi-classical spectrum  should ``Reggeize" to
$M^2_{nl}\approx 2\pi \sigma_T (n+l)$ for the $n$-radial and $l$-orbital excitations.

Another issue is that $M^2$ appears not only in the l.h.s. of (\ref{LIGHT3}) but also in the r.h.s.
For $heavy$ mesons on the light cone, one can assume $M\approx 2m_Q$  on the right-hand-side 
\bea
\label{HEAVY}
M_H^2&\approx &
\frac {m_Q^2+k_\perp^2}{x\bar x}+2\sigma_T\bigg(|id/dx|^2+(2m_Q)^2 x_\perp^2\bigg)^{\frac 12}\nonumber\\
\eea
and avoid the iterative process.
Furthermore,  in the heavy-quark limit, $x\approx \bar x\approx \frac 12$  and (\ref{HEAVY}) simplifies to
\bea
\label{HEAVYX}
M_H^2&\approx &
(2m_Q)^2+4k_\perp^2+4 m_Q\sigma_T |x_\perp|
\eea
Note that the effective string tension is now growing with $m_Q$. This is reasonable, since the binding energy for a slowly moving
but heavy quark, should compensate its kinetic energy $m_Qv_Q^2/2$, which depends on $m_Q$. The semi-classical spectrum
follows from

\begin{widetext}
\be
\int_{\rho_-}^{\rho_+}\,d\rho \bigg(M_{H,nl}^2-(2m_Q)^2-\frac {4l^2}{\rho^2}-4m_Q\sigma_T \rho\bigg)^{\frac 12}=2\pi n
\ee
\end{widetext}
with the turning points $\rho_\pm$ fixed by the positive and real solutions to the cubic equation

\be
M_{H,nl}^2=(2m_Q)^2+\frac {4l^2}{\rho_\pm^2}+4m_Q\sigma_T \rho_\pm
\ee
which exist for

\be
M_H^2\geq (2m_Q)^2+8(l m_Q\sigma_T)^{\frac 23}
\ee
For $l=0$, the radial excitations follow 

\be
M_{H,n0}^2\approx (2m_Q)^2+\bigg(\frac{6m_Q}{\alpha^\prime}\bigg)^{\frac 23}\,n^{\frac 23}
\ee
with again, the open string Regge slope  $\alpha^\prime=1/2\pi\sigma_T$. As we noted earlier, the Regge trajectory is now bent.
The heavier $m_Q$, the more bound  the Regge spectrum.
(\ref{HEAVY}) applies also to  heavy-light mesons modulo minor changes for asymmetric masses.

Returning to light-light mesons, we note that the nonrelativistic approximation $M\approx 2m_Q$ may still be semi-quantitatively suited for the ``ordinary mesons" (like
vectors $\phi,\rho$) but not the Nambu-Golstone pseudoscalars.
In order to reproduce the mesonic spectra correctly,  (\ref{LIGHT3}) needs to be
supplemented by the  spin and flavor-dependent interactions, as we will discuss later. 

Finally, 
we note that if the constituent mass $m_Q\rightarrow 0$,  (\ref{LIGHT3}) admits a non-normalizable massless solution $\varphi_0(x, b_\perp) \sim \theta (x\bar x)$ for $x\bar x\neq 0$ since

\bea
\label{LIGHT4}
&&\bigg[M^2\bigg]_{m_Q=0}\varphi_0(x,k_\perp)=\nonumber\\
&&\bigg[\frac {k_\perp^2}{x\bar x}+2\sigma_T |id/dx|\bigg]\theta(x\bar x)=0
\eea
At  $x=0,1$ it  vanishes  as  a power proportional to the mass $(x\bar x)^{\# m_Q}$. This is the same massless solution in the 1+1 dimensional $^\prime$t Hooft 
equation (\ref{THOOFT}) as the would-be-pion emerges as a massless mode
in the chiral limit at large $N_c$ only if $m_Q=0$ is  identified with the current quark mass. However, this kinematical
solution is not the physical pion, since for the latter $m_Q\neq 0$ is identified with the constituant mass, and does not vanish in the chiral limit.

\subsection{Eliminating the square root in the Hamiltonian}


 The inconvenient square root of the differential operator can be avoided by the ``einbein trick". With this in mind, consider the operator

\bea
\label{LIGHT4}
M^2(a,b )=
\frac {m_Q^2+k_\perp^2}{x\bar x}+\sigma_T\bigg(\frac{|id/dx|^2+b x_\perp^2}a+a\bigg)\nonumber\\
\eea
with the auxillary parameters $a$ (inverse einbein $a=1/e$) and $b$. Note that minimization over $a$ would
return us to the original Hamiltonian with the square root: but we will do minimization is $a$ $afterwords$, after diagonalization. The
parameter $b$ would be iteratively selected to reach self-consistency when $b\rightarrow M_{nl}^2$.
(Strictly speaking $b$ is originally $M^2$, so  the minimization in $a$ is  more subtle. This  subtlety will
be ignored for now, as well as the ordering issue we pointed out).

Let us represent this
  Hamiltonian as a sum of two terms, $M^2\equiv H_0+V$,
\begin{equation}
 H_0={\sigma_T \over a} \bigg( -{\partial^2  \over \partial x^2}-b{\partial^2  \over \partial  k_\perp^2} \bigg) + \sigma_T  a + 4(m_Q^2+ k_\perp^2)
\end{equation} 
Note that the last term is artificially added: it is then subtracted from  the ``potential" defined by
\begin{equation}  
V(x,\vec k_\perp)\equiv (m_Q^2+k_\perp^2)\bigg({1 \over x \bar x} -4\bigg)
\end{equation} 
 $H_0$ selects the functional basis set 
 described in Appendix~\ref{basis}.   
 The Hamiltonian  $M^2$ consists of the diagonal part $H_0$, and non-diagonal 
 ``potential" $V$ part. In the orthonormal  set of functions defined in Appendix~\ref{basis},  $M^2$ is represented by (infinite) matrices,
 with its $12\times 12$ part given explicitly. 
 
\begin{figure}[h!]
	\centering
		\includegraphics[width=7cm]{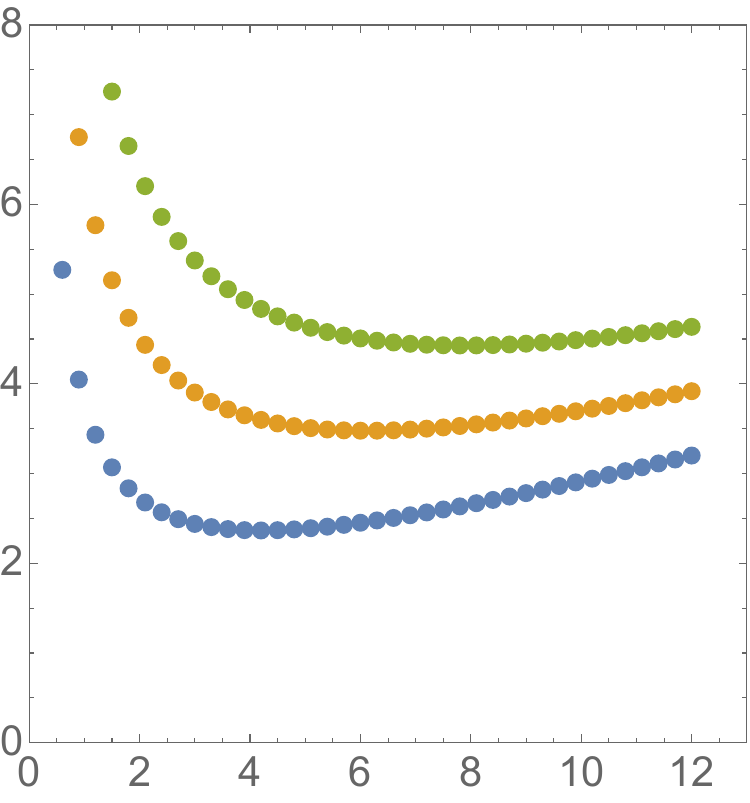}\\
	\includegraphics[width=7cm]{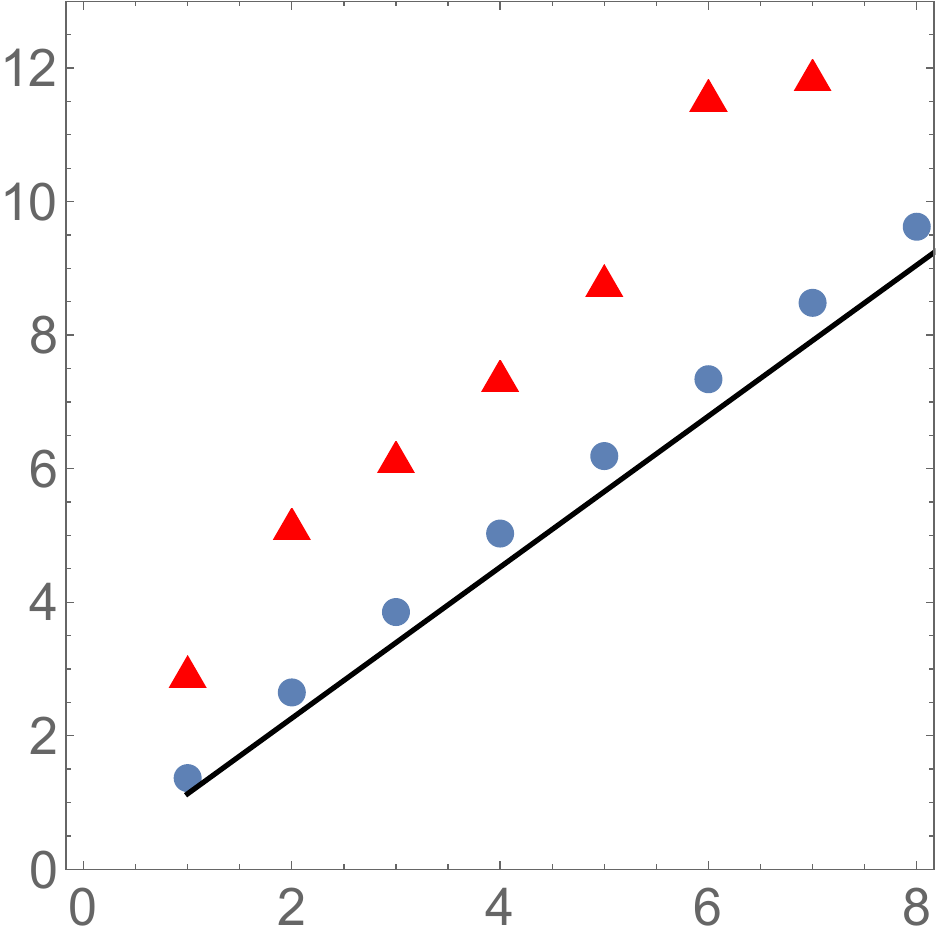}
	\caption{The upper plot shows the dependence of the three lowest eigenvalues $M^2_n$ on the parameter $a$,
	in  the region around their mimima. The lower plot shows  $M^2_{n+1}$ versus $n+1=1..7$.
	The results obtained from the Hamiltonian diagonalization with fixed $a=2.36$, are shown by red triangles. The blue disks are  the
	semiclassical results discussed previously, the line is a simple linear expression $M^2_n=n/\alpha'$, both shown for comparison. }
	\label{fig:comparem2}
\end{figure}

 For the simple ``bare mass approximation"  with the parameter $b=(2 m_Q)^2$, one
 can diagonalize the (part of the) Hamiltonian. The dependence of the (three lowest) eigenvalues on 
 the parameter $a$ is shown in the upper plot of Fig.~\ref{fig:comparem2}.  There are clear minima as a function of $a$. While they
 do not happen to be at the same values, the dependence  $M^2_n(a)$ is relatively minor, and selecting a certain compromise value 
 gives reasonable numerical accuracy. We use $a=2.36$ (a minimum for the lowest $n=1$ state).
 
 With the  parameters $a,b$  fixed, the  Hamiltonian becomes a numerical matrix, which can be readily diagonalized. 
 Keeping the 12$\times$12 part of the matrix, we  obtain the 12 eigenvalues 
 shown   in the lower part of  Fig.~\ref{fig:comparem2} (red  triangles). For comparison we also show
 the semiclassical results. The Regge slope is well reproduced, while the
 intercept $a_M$ is not. This can be attributed to the fact that our LF Hamiltonian includes zero
 mode oscillation energy in all three directions, missing in  the  semiclassical treatment.

In principle, 
 one may tune the parameters $a,b$ for each state separately,
to reach agreement for the masses. However, this would mean
 that different states are not eigenstates of the same Hamiltonian, and therefore not mutually orthogonal.
Instead of doing that, we keep the $a,b$ values the same for all considered states, and look at the main object of our interest, the derived wave functions. For example, in the approximation considered,  the lowest state has the
following LFWF
\begin{widetext}
\begin{eqnarray} \label{LFWF_1}
\psi_1(\rho, x)=&& \beta \,e^{- \beta^2 \rho^2/2}\,\bigg((0.831 - 
0.0371 \beta^2 \rho^2 + 
0.00100 \beta^4 \rho^4)\, {\rm sin}(\pi x) \nonumber \\
&&+ (-0.0252 - 
0.0107 \beta^2 \rho^2 + 
0.000566 \beta^4 \rho^4) \, {\rm sin}(3 \pi x) \\
&&+(- 
0.00427  - 
0.00207 \beta^2 \rho^2 + 
0.000168 \beta^4 \rho^4) \, {\rm sin}(5 \pi x) \nonumber \\ 
&&+(- 
0.00145  - 
0.000743 \beta^2 \rho^2  + 
0.0000633 \beta^4 \rho^4) \, {\rm sin}(7 \pi x)\bigg)\nonumber
\end{eqnarray}
\end{widetext}
We recall that here  $\rho=p_\perp$,  and the oscillator parameter is $\beta=(4a/\sigma_T b)^{\frac 14}$.

Note that only the first coefficient (0.831) is large, while the 
others are at few percent level or smaller. If plotted, 
the Gaussian curve is hard to separate from the full expression.
This means that
the main $p_\perp$ dependence is mostly Gaussian, while the $x$ dependence is nearly $\sim {\rm sin}(\pi x)$
(amusingly as in 1+1-dimensions). 
This is explained by  our definition of
the oscillator term $4(m_Q^2+p_\perp^2)$ (which was added and subtracted), making the non-diagonal matrix elements of the Hamitonian  relatively small. 
Using the Gaussian approximation for the ground state,  the absolute scale of the r.m.s. of the transverse momentum is

\begin{equation}
\langle p_\perp^2 \rangle={1\over \beta^2}=\sqrt{{
	  \sigma_Tb  \over	4 a  }}
\end{equation}

Finally, (\ref{LIGHT4}) can be solved  in three dimensions $x, p_x, p_y$ without recourse
to the matrix diagonalization. The method consists of solving directly the partial differential equation,
with Dirichlet boundary conditions on the support with Bjorken-x. 
The ground state eigenvalue is $M^2=2.58\, GeV^2$, in good agreement with the matrix diagonalization. 
The difference between the
numerical solution and (\ref{LFWF_1}) is inside the width of the line. No change in the normalization was needed.
Orthogonality also works very accurately.

\begin{figure}[htbp]
\begin{center}
\includegraphics[width=6cm]{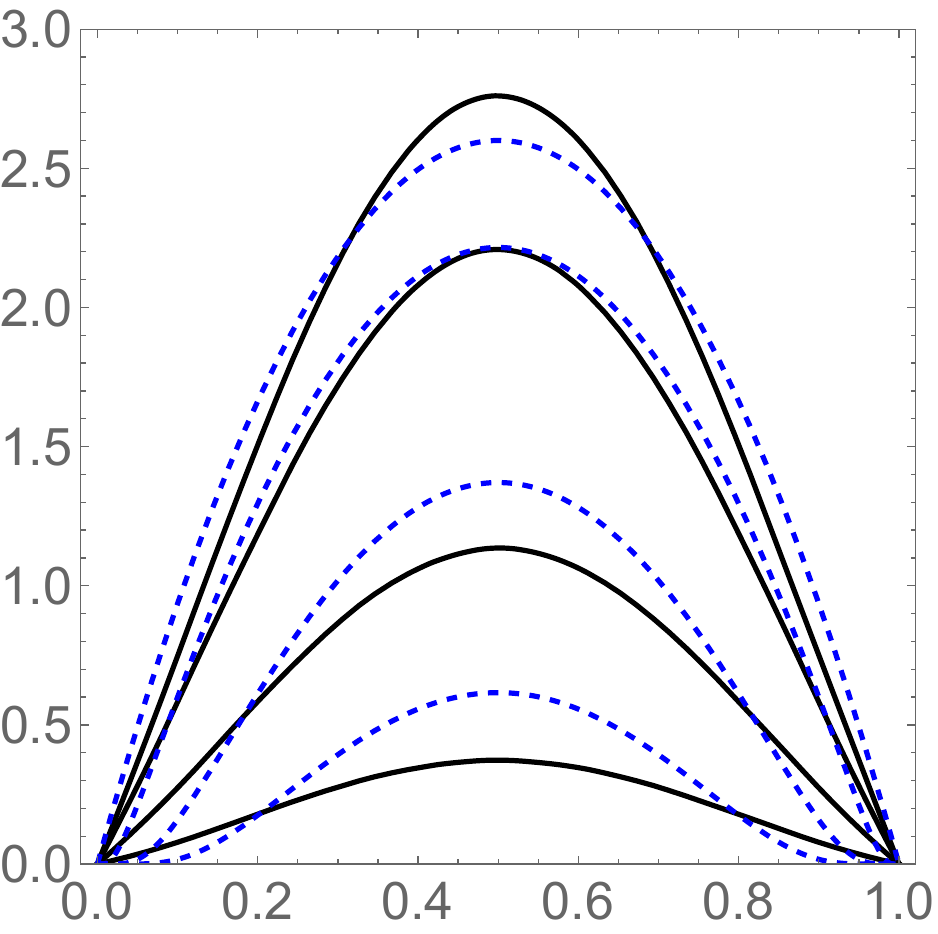}
\includegraphics[width=6cm]{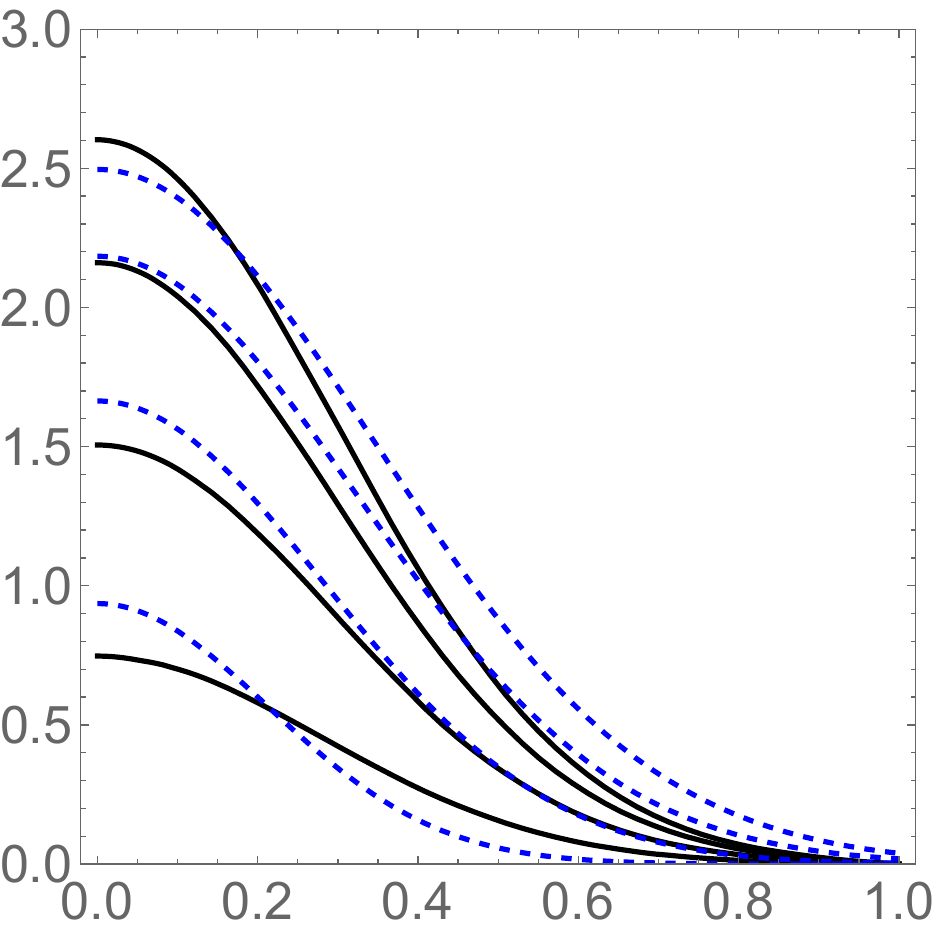}
\caption{The upper plot shows the dependence of the ground state wave function $\Psi_0(x,p_\perp)$ versus
$x$ ar $p_\perp^2=0,0.2,0.4,0.6 \, GeV^2$, top to bottom curves. The lower plot is versus $p_\perp$ at  $x=0.1,0,2,0.3,0,4$. All solid curves are from the
exact solution, while all dashed lines are for  the simplified  form (\ref{eqn_fit}). }
\label{fig_ground}
\end{center}
\end{figure}

Since we have two solutions, one numerical and one analytical  (\ref{LFWF_1}), they can be plotted in various ways.
One way, is to make a comparison to the form suggested in the litterature
\be  
\label{FIT}
\Psi_0(x, \xi)=4C_2\,x \bar x\,e^{-C_1 \xi^2}\,,
\ee
with 
\be
\label{XIBT}
\xi^2={p_\perp^2 \over x \bar x}
\ee
 the Brodsky-DeTeramond variable. The constants are fixed as 
$C_2=2.6,C_1=1$. The compartison between our exact ground state (\ref{LFWF_1}) and (\ref{FIT}), is shown in~Fig.\ref{fig_ground}.
The  simplified form (\ref{FIT}) with the $\xi$ variable, is qualitatively similar but not very accurate, especially at the end points.

%
%
%

\section{Other light front Hamiltonians}
\label{sec_other_H}

The method of writing a Hamiltonian as a large matrix in some convenient basis functions, with its subsequent
diagonalization, is widely used in atomic and nuclear physics.
 Jia and Vary in~\cite{Jia:2018ary}  pioneered such approach to LFWFs.
 Their  assumed Hamiltonian consists of 
four  terms (i) the effective quark masses originating from the spontaneous breaking of chiral symmetry ($H_M$); (ii) the longitudinal confinement ($H_{||}$);
(iii) the transverse motion and confinement ($H_\perp$);   and last but not least, (iv) the NJL 4-quark effective interaction $H_{NJL}$ we will not detail here. More specifically,

\bea
H&=&H_M+H_{||}+ H_\perp+H_{NJL} \nonumber\\
H_M&=&{m_Q^2 \over x_1}+{m_{\bar Q}^2 \over x_2} \nonumber\\
H_{||}&=&{\kappa^4 \over (m_Q+m_{\bar Q})^2} {1 \over J(x) }\partial_x J(x) \partial_x\nonumber\\
H_\perp&=&k_\perp^2 \bigg({1 \over x_1}+{1 \over x_2}\bigg) +\kappa^4 x_1x_2 r_\perp^2 \nonumber\\
\eea
where $m_{Q,\bar Q}$ are the constituent   quark and antiquark masses, $\kappa$ is the confining parameter, $J(x)=x_1x_2=(1-s)^2/4$
is the integration measure, and $\vec k_\perp,\vec r_\perp$ are the transverse momentum and coordinate. 
Note that if the masses are the same, 
one can simplify $${1\over x_1}+{1 \over x_2}={1 \over x_1x_2}={4 \over (1-s^2)}  $$
The matrix element of $H_M$  lacks the factor  of $(1-s^2)$ normally present in their integration measure.

Note that the confining terms are quadratic (rather than linear) in coordinates, and the transverse and longitudinal parts are additive. Therefore it resembles harmonic oscillators. This  simplifies the problem of finding its eigenfunctions. Those are explicitly defined in   \cite{Jia:2018ary}. We  will
not explain them here, but just mention that with those the Hamiltonian 
(other than NJL) is diagonal, and  its spectrum is analytic 
\bea 
M^2_{nml}&=&(m_Q+m_{\bar Q})^2+ 2\kappa^2 (2n+ |m|+l_L+3/2) \nonumber\\
&+&{\kappa^4 \over (m_Q+m_{\bar Q})^2}l_L(l_L+1)
\label{eqn_JV_spectrum}
\eea
where $n$ is the principal quantum number of the transverse oscillator, $m$ is the angular momentum helicity
(from $exp(i m \phi)$), and $l_L$ is the index of the longitudinal wave functions (not orbital momentum).

The  term $k^2_\perp/x\bar{x}$ is kinematically natural
on the light front.  As it depends both on the transverse and longitudinal momenta -- which we try to separate -- by
using a certain expansion in basis functions,  in the diagonalization presented above.

Trying to cut through these difficulties (and following Brodsky et al.) Jia and Vary proposed
 to  change variables  to $\xi$ as given  in (\ref{XIBT}), 
and
call the offensive term its square. Unfortunately, if this change of variables is to be done explicitly,
 it would complicate significantly the ``kinetic energy" of the problem, producing
extra terms which were not included. 


Let us now compare this spectrum with that of our ``basic problem" discussed above. 
Note first, that the main linear dependence on the integer quantum numbers $n,m$ is
in agreement with   the linear Regge trajectories. So  qualitatively it is in agreement with the data.

Unfortunately, the particular selection of the parameter $\kappa=0.227\, GeV$ makes the slope
of the resulting Regge trajectory much
smaller than needed: see the open points in~Fig.\ref{fig:basicm2} (for m=l=0).  To fix this, 
one needs a larger value, such as e.g. $\kappa=0.5\, GeV$ as indicated by the dashed line
on the same figure.

Another test, using the same Fig.~\ref{fig:basicm2}, can be made using the $\omega_3-\omega$
spliting, corresponding to a change in $m$ by two units. From the expression above, one finds that
it should be $4\kappa^2\approx 0.2\, GeV^2$ if the recommended value $\kappa=0.227\, GeV$
is used. Experimentally, for the three lowest $\omega_3,\omega$ states it is $\approx 1.8\, GeV^2$, nearly an order of magnitude larger. 

We conclude that while the description of confinement by Jia-Vary Hamiltonian
is qualitatively correct, leading 
to a Regge-type behavior, the particularly recommended value of the parameter $\kappa$ leads
to a significant underestimation of the confinement effects. The reason is  that their analysis
focused on the lowest states -- specifically on $\pi,\rho$ mesons  -- rather than on Regge phenomenology of
the excited states. However (as we detailed in our paper~\cite{I}) $\rho$ and especially $\pi$ mesons are 
very special case.  Being most compact in size,
they are strongly  affected by the short-range effects (spin potentials and 
residual interactions), rather than  generic effects of confinement.

\section{Central potential from instantons on the Light Front}
\label{lwilson}

In the first paper of this series \cite{I} we discussed a  ``dense instanton liquid" model,
including both the dilute instanton ensemble of the original ILM, responsible for  the
disordering of the lowest Dirac eigenstates, as well as the  ``$I\bar I$ molecules" with a larger density.
We have shown that such a vacuum model can
reproduce $both$ the $central$ potential $V_C(r)$ at intermediate distances $r\sim 0.5\, fm$, and 
the nonperturbative {\em spin-dependent } forces. We recall that the 
use of instantons  is motivated by the fact that the spin-dependent  forces
stem from (a nonlocal correlator of) $magnetic $ fields.

To evaluate the light cone Wilson loop in Fig.~\ref{fig_qqbar} in the instanton vacuum, we follow our original idea for $Q\bar Q$ scattering in~\cite{Shuryak:2000df}.
For that we assign a relative angle $\theta$ between Wilson lines in Euclidean space, carry the summation and tracing over the ensemble of instantons 
with fixed density $N/V_4$, and then analytically continue the $result$ to the light front using the substitution $\theta=-i\chi$ with $\chi$ being the rapidity 
difference of the $Q\bar Q$ beams. The  connected loop can be written in terms of traces over individual instantons in leading order in $N/V_4$. 
More explicitly, the connected result exponentiates to

\begin{widetext}
\bea
\label{W1}
\langle {\bf W}(\theta, 0_\perp)\,{\bf W}^\dagger(\theta, b_\perp)\rangle_C\approx 
{\rm exp}\bigg(-2\times \frac{N}{2N_cV_4}\int d^4z\,{\rm Tr}_c\bigg({\bf 1}-{\bf W}_I(\theta, 0_\perp){\bf W}_I^\dagger(\theta,b_\perp)\bigg) \bigg)
\nonumber\\
\eea
\end{widetext}
with ${\bf W}_I(\theta, b_\perp)$ the sloped Wilson line running through an instanton at a transverse separation $b_\perp$.
The extra overall factor of 2 in the exponent accounts for  the anti-instanton contribution.

\subsection{Case $\dot{x}\cdot b_\perp=0$}

Each sloped Wilson line contributes in singular gauge

\bea
\label{W2}
{\bf W}_I(\theta,b_\perp)=
{\rm cos}\bigg(\pi-\frac{\pi\gamma}{\sqrt{\gamma^2+\rho^2}}\bigg) \\
-i\hat{n}^a\tau^a{\rm sin}\bigg(\pi-\frac{\pi\gamma}{\sqrt{\gamma^2+\rho^2}}\bigg)
\nonumber
\eea
with 

\bea
\label{W3}
n^a=&&\eta^a_{\mu\nu}\dot{x}_\mu (z-b)_\nu\\
 \gamma^2=&&n\cdot n= (z_4{\rm sin}\theta-z_3{\rm cos}\theta)^2+(z_\perp-b_\perp)^2 \nonumber 
\eea
In this first case $b_\mu=(0, b_\perp, 0)$, $x_\mu(s)=({\rm cos}\theta s, 0_\perp, {\rm sin}\theta s)$ with $\dot{x}\cdot b_\perp=0$.
Carrying the color trace and using the new coordination

\bea
\label{W4}
z_-&=&{\rm sin}\,\theta z_4-{\rm cos}\,\theta z_3\nonumber\\
z_+&=&{\rm cos}\,\theta z_4+{\rm sin}\,\theta z_3
\eea
yield the result

\begin{widetext}
\bea
\label{W5}
\frac{2N}{N_c V_4}\int dz_+dz_-dz_\perp 
\bigg[1-{\rm cos}\bigg( \frac{\pi\tilde\gamma}{\sqrt{\tilde\gamma^2+\rho^2}}\bigg)
{\rm cos}\bigg( \frac{\pi\tilde{\underline\gamma}}{\sqrt{\tilde{\underline\gamma}^2+\rho^2}}\bigg)\nonumber\\
-\frac{z\cdot b_\perp -z_-^2-z_\perp^2}{\tilde\gamma\tilde{\underline\gamma}}
{\rm sin}\bigg( \frac{\pi\tilde\gamma}{\sqrt{\tilde\gamma^2+\rho^2}}\bigg)
{\rm sin}\bigg( \frac{\pi\tilde{\underline\gamma}}{\sqrt{\tilde{\underline\gamma}^2+\rho^2}}\bigg)\bigg]
\eea
with

\bea
\label{W6}
{\tilde\gamma}^2=z_-^2+(z_\perp-b_\perp)^2\qquad
{\tilde{\underline\gamma}}^2=z_-^2+z_\perp^2
\eea
Since (\ref{W5}-\ref{W6}) are $z_+$ independent,  the result scales with $Z^+_E=\int dz_+$

\bea
\label{W7}
\langle {\bf W}(\theta, 0_\perp)\,{\bf W}^\dagger(\theta, b_\perp)\rangle_C\approx 
{\rm exp}\bigg[-Z_E^+\bigg(\frac{4\kappa}{N_c\rho}\bigg)\,{\bf I}\bigg(\xi=\frac{b_\perp}\rho\bigg)\bigg]
\eea
and the dimensionless cylindrical integral

\bea
\label{W8}
{\bf I}(\xi )=&&\int_{-\infty}^{+\infty} dy_{-}\int_0^\infty y_\perp dy_\perp\int_0^{2\pi} \frac{d\phi}{2\pi} \nonumber\\
&&\times\bigg[1-{\rm cos}\bigg(\frac{\pi y}{\sqrt{y^2+1}}\bigg)
{\rm cos}\bigg(\pi\bigg(\frac{y^2+\xi^2+2\xi y_\perp \,{\rm cos}\phi}{y^2+\xi^2+2\xi y_\perp \,{\rm cos}\phi+1}\bigg)^{\frac 12}\bigg)\nonumber\\
&&-\frac{y^2+\xi  y_\perp{\rm cos}\phi}
{(y^2(y^2+\xi^2+2\xi  y_\perp {\rm cos}\phi))^{\frac 12}}
{\rm sin}\bigg(\frac{\pi y}{\sqrt{y^2+1}}\bigg)
{\rm sin}\bigg(\pi\bigg(\frac{y^2+\xi^2+2\xi y_\perp \,{\rm cos}\phi}{y^2+\xi^2+2\xi  y_\perp \,{\rm cos}\phi+1}\bigg)^{\frac 12}\bigg)\bigg]\nonumber\\
\eea
\end{widetext}
with the radial  variable   $y^2=y_-^2+y_\perp^2$.  Note that for the temporal Wilson loop or $\theta=0$ in our cae, a similar integral arises for the
static potential between two infinitly heavy $Q\bar Q$ with two major differences: 1/ the dimensionless integral involves spherical coordination
and a spherical measure; 2/ $y_\perp\,{\rm cos}\phi \rightarrow y\,{\rm cos}\theta_S$ with $\theta_S$ the spherical angle.

Although $\theta$ has dropped out of (\ref{W7})  it is worth noting through (\ref{W4}) that $Z^+_E$ analytically continues to the transverse light cone coordinate $iZ_M^+$.
With this in mind, (\ref{W7}) analytically continues to 

\begin{widetext}
\bea
\label{W10}
\langle {\bf W}(\theta, 0_\perp)\,{\bf W}^\dagger(\theta, b_\perp)\rangle_C\rightarrow 
{\rm exp}\bigg[-i\,Z_M^+\,\bigg(\frac{4\kappa}{N_c\rho}\bigg)\,{\bf I}\bigg(\frac{b_\perp}\rho\bigg)\bigg]
\eea
\end{widetext}
which allows for the identification of the instanton contribution to the light cone Hamiltonian

\bea
\label{11}
P_I^- =\frac 1{\gamma_\beta}\bigg(\frac{4\kappa}{N_c\rho}\bigg)\,{\bf I}\bigg(\frac{b_\perp}\rho\bigg)
\eea
with the extra Lorentz factor $\gamma_\beta={\rm cosh}\chi$ correcting for the missing time-dilatation factor  in the exponent in (\ref{W10}).
The corresponding  instanton contribution to the invariant squared mass is

\bea
\label{12}
2P^+P^-_I=2\gamma_\beta M\,P_I^-\approx 2M \bigg(\frac{4\kappa}{N_c\rho}\bigg)\,{\bf I}\bigg(\frac{b_\perp}\rho\bigg) \nonumber\\
\eea
in leading order in the packing  fraction $\kappa$.
In the chiral limit with zero current quark masses, the full squared mass (without the confining string) is kinetic plus potential

\be
\label{13}
M^2=\frac{k_\perp^2}{x\bar x}+2P^+P^-_I\approx \frac{k_\perp^2}{x\bar x}+2M \bigg(\frac{4\kappa}{N_c\rho}\bigg)\,{\bf I}\bigg(\frac{b_\perp}\rho\bigg) 
\ee
which amounts to the mass operator

\be
\label{14}
M=\frac{|k_\perp |}{\sqrt{x\bar x}}+\bigg(\frac{4\kappa}{N_c\rho}\bigg)\,{\bf I}\bigg(\frac{b_\perp}\rho\bigg) +{\cal O}(\kappa^2)
\ee
For $\xi=b_\perp/\rho \ll 1$ the transverse potential is harmonic with  ${\bf I}(\xi)\approx \alpha \xi^2$, while for $\xi=b_\perp/\rho \gg 1$ the transverse
potential asymptotes twice the induced self-energy ${\bf I}(\xi)\approx 2\Delta m_Q+C/\xi^p$ with $p\ll 1$. Typically, the self energies on the Wilson lines 
are small $\Delta m_Q/m_Q<1$.
\\
\\
\subsection{Case $\dot{x}\cdot b_\perp\neq 0$}

In this second case $b_\mu=(0, b_\perp, b_3)$, $x_\mu(s)=({\rm cos}\theta s, 0_\perp, {\rm sin}\theta s)$ with $\dot{x}\cdot b_\perp\neq 0$
The analysis follows the same reasoning without a longitudinal component $b_3$, with 
(\ref{W5}) now reading

\begin{widetext}
\bea
\label{W5X}
&&\frac{2N}{N_c V_4}\int dz_+dz_-dz_\perp 
\bigg[1-{\rm cos}\bigg( \frac{\pi\tilde\gamma}{\sqrt{\tilde\gamma^2+\rho^2}}\bigg)
{\rm cos}\bigg( \frac{\pi\tilde{\underline\gamma}}{\sqrt{\tilde{\underline\gamma}^2+\rho^2}}\bigg)\nonumber\\
&&-\frac{z_-^2+{\rm cos}\theta\,z_-b_3+z_\perp\cdot (z-b)_\perp}{\tilde\gamma\tilde{\underline\gamma}}
{\rm sin}\bigg( \frac{\pi\tilde\gamma}{\sqrt{\tilde\gamma^2+\rho^2}}\bigg)
{\rm sin}\bigg( \frac{\pi\tilde{\underline\gamma}}{\sqrt{\tilde{\underline\gamma}^2+\rho^2}}\bigg)\bigg]
\eea
\end{widetext}
and 

\bea
\label{W6X}
{\tilde\gamma}^2&=&(z_-+{\rm cos}\theta \,b_3)^2+(z_\perp-b_\perp)^2\nonumber\\
{\tilde{\underline\gamma}}^2&=&z_-^2+z_\perp^2
\eea
We now  analytically continue $\theta\rightarrow -i\chi$ or $\rm{cos}\theta\rightarrow {\rm cosh}\chi=\gamma_\beta$, and $Z_E^+\rightarrow iZ_M^+$, and  change  to the dimensionless variables
$z_-/\rho\rightarrow z_-$ and $z_\perp/\rho  \rightarrow z_\perp$. The result for (\ref{W5X}) is now

\bea
\label{W7X}
iZ^+_M\,\frac{2N\rho^3}{N_c V_4}\,{\bf H}\bigg(\frac 1{M\rho}\frac{id}{dx}, \frac{b_\perp}{\rho}\bigg)
\eea
with the dimensionless integral

\begin{widetext}
\bea
\label{W8X}
&&{\bf H}\bigg(\frac 1{M\rho}\frac{id}{dx}, \xi\bigg)=\frac 1{2\pi}
\int dz_-dz_\perp 
\bigg[1-{\rm cos}\bigg( \frac{\pi\tilde\gamma}{\sqrt{\tilde\gamma^2+1}}\bigg)
{\rm cos}\bigg( \frac{\pi\tilde{\underline\gamma}}{\sqrt{\tilde{\underline\gamma}^2+1}}\bigg)\nonumber\\
&&-\frac{z_-(z_-+id/dx/M\rho)+z_\perp\cdot (z_\perp-\xi_\perp)}{\tilde\gamma\tilde{\underline\gamma}}
{\rm sin}\bigg( \frac{\pi\tilde\gamma}{\sqrt{\tilde\gamma^2+1}}\bigg)
{\rm sin}\bigg( \frac{\pi\tilde{\underline\gamma}}{\sqrt{\tilde{\underline\gamma}^2+1}}\bigg)\bigg]
\eea
with $\xi_\perp=b_\perp/\rho$ and $\xi=|\xi_\perp|$ and 

\bea
\label{W9X}
{\tilde\gamma}^2&\rightarrow& (z_-+id/dx/M\rho)^2+(z_\perp-\xi_\perp)^2\nonumber\\
{\tilde{\underline\gamma}}^2&\rightarrow& z_-^2+z_\perp^2
\eea
The integral in (\ref{W8X}) is only a function of the combination 

\be
\label{W99X}
\tilde\xi_{x}=((id/dx/M\rho)^2+\xi_\perp^2)^{\frac 12}\equiv\xi_x/\rho
\ee
 with the spherical integral

\bea
\label{W10X}
{\bf H}(\xi_{x} )=&&\int_0^\infty y^2 dy\int_{-1}^{+1} dt \nonumber\\
&&\times\bigg[1-{\rm cos}\bigg(\frac{\pi y}{\sqrt{y^2+1}}\bigg)
{\rm cos}\bigg(\pi\bigg(\frac{y^2+\tilde\xi_x^2+2\tilde\xi_{x}yt}{y^2+\tilde\xi_x^2+2\tilde\xi_{x}yt+1}\bigg)^{\frac 12}\bigg)\nonumber\\
&&-\frac{y+\tilde\xi_{x}t}
{(y^2+\tilde\xi_x^2+2\tilde\xi_{x}yt)^{\frac 12}}
{\rm sin}\bigg(\frac{\pi y}{\sqrt{y^2+1}}\bigg)
{\rm sin}\bigg(\pi\bigg(\frac{y^2+\tilde\xi_x^2+2\tilde\xi_{x}yt}{y^2+\tilde\xi_x^2+2\tilde\xi_{x}yt+1}\bigg)^{\frac 12}\bigg)\bigg]
\eea

The corresponding instanton  contribution to the invariant squared mass is  now


\be
\label{16Z} 
M^2\approx  \frac{k_\perp^2+m_Q^2}{{x\bar x}} + 2P^+P_I^-\approx 
 \frac{k_\perp^2+m_Q^2}{{x\bar x}}+ 2M \bigg(\frac{4\kappa }{N_c \rho}\bigg){\bf H}(\tilde\xi_x)\equiv 
  \frac{k_\perp^2+m_Q^2}{{x\bar x}}+ 2M V_{C}(\xi_x)
\ee
which is an iterative equation for the mass $M$. ${\bf H}(\xi_x)$ admits the short and large distance limits

\bea
\label{17Z} 
{\bf H}(\tilde\xi_x) \approx&&+ \bigg(\frac{\pi^3}{48}-\frac{\pi^3}3J_1(2\pi )\bigg)\tilde\xi_x^2+
\bigg(-\frac{\pi^3(438+7\pi^2)}{30720}+\frac{J_2(2\pi)}{80}\bigg)\tilde\xi_x^4\nonumber\\
{\bf H}(\tilde\xi_x) \approx &&-\frac{2\pi^2}3\bigg(\pi J_0(\pi)+J_1(\pi)\bigg)+\frac{C}{\tilde\xi^p_x}
\eea
\end{widetext}

\section{Spin interactions on the light front}
\label{SPIN-NZM}

To construct the spin-dependent interactions  
 on the light front,  we apply the general construction by Eichten and Feinberg~\cite{Eichten:1980mw},
to the slated Wilson loop shown in  Fig.~\ref{fig_qqbar} in Euclidean signature, followed by the analytical continuation $\theta\rightarrow -i\chi$ to Minkowski signature. For that, we frst need the expansion of the heavy-quark propagator shown as a straight line in leading order or $1/m_Q^0$,  
at next-to-next to leading order.

\subsection{Heavy-quark reduction}

The heavy quark expansion of a Dirac fermion of mass $m_Q$ with fixed velocity, 
in an arbitrary gauge field is best achieved using the Foldy-Wuthuysen transformation on the relativistic
fermion propagator, 

\be
\label{EPOT}
e^{-i\frac{\slashed{D}_\perp}{2m_Q}}\,\frac 1{i\slashed{D}-m_Q}\,e^{-i\frac{\slashed{D}_\perp}{2m_Q}}
\ee
with $i\slashed{D} =i\slashed{\partial}+\slashed{A}$ and $\slashed{D}_\perp =\slashed{D}-\slashed{v}v\cdot D$ satisfying $[\slashed{D}_\perp, \slashed{v}]_+=0$.
We will refer to $v_\mu$ the 2-dimensional {\it light-cone-like} velocity along the 2-dimensional {\it light-cone-like} coordinate  $x_+$ in Euclidean signature, 
and to $v_{\perp \mu}$ its orthogonal  velocity along the 2-dimensional {\it light-cone-like} coordinate $x_-$ also in Euclidean signature,

\bea
&&v_\mu=({\bf 0}_\perp,  {\rm sin}\theta,{\rm cos}\theta),\nonumber\\
&&v_{\perp \mu}=({\bf 0}_\perp,  -{\rm cos}\theta,{\rm sin}\theta),
\eea
with $x_+=v\cdot x$ and $x_-=v_\perp\cdot x$.
These light-cone-like Euclidean coordinations (lower indices) are not to be confused with the Minkowski light-cone coordinates $x^\pm=x^0\pm x^3$
(upper-indices).  With this in mind, and to order $1/m_Q^2$ the heavy quark propagator is

\begin{widetext}
\begin{eqnarray}
\label{NR}
\frac 1{iv\cdot D}-\frac 1{iv\cdot D}\bigg(\frac 1{2m_Q}(i\slashed D_\perp)^2
-\frac 1{4m^2_Q}(i\slashed D_\perp)(iv\cdot D)(i\slashed D_\perp)\bigg) \frac 1{iv\cdot D} 
\end{eqnarray}
The bracket in (\ref{NR}) gives rise to a vertex insertion, which can be re-arranged

\be
\label{NR2}
\frac 1{2m_Q}\bigg((iD)^2-\frac 12\sigma_{\mu\nu}F_{\mu\nu}\bigg)
-\frac 1{4m^2_Q}\bigg(i\sigma_{\alpha\nu}iD_{\alpha} v_\mu  F_{\mu\nu}+ iD_{\nu} v_\mu F_{\mu\nu}\bigg)
\ee
\end{widetext}
with $\sigma_{\alpha\nu}=\frac 1{2i}[\gamma_\alpha, \gamma_\nu]$.  In (\ref{NR2})
we have dropped all terms that vanish on-shell, i.e $v\cdot D Q_v=0$
with $Q_v$ the heavy quark field.
When  inserted on a straight Wilson line, (\ref{NR2})  produces the spin corrections up to order $1/m_Q^2$.  

Note that in the Dirac representation $\sigma_{4i}$ is off-diagonal. The electric contribution  mixes  particles and anti-particles. It  does not contribute 
when inserted on a straight Wilson line defined as

\bea
{\bf W}(y,x)=&&\langle y_+|\frac 1{v\cdot D}|x_+\rangle \,\delta({\bf{x}}-{\bf{y}})\nonumber\\
=&&{\bf P}e^{i\int_{x_+}^{y_+}A\cdot dz}\,\theta(y_+-x_+)\delta({\bf{x}}-{\bf{y}})\nonumber\\
\eea
with the ordering along $x_+$ and the short hand notations 

\bea
&&x_\mu=({\bf x}_\perp, x_-,x_+)\equiv ({\bf{x}}, x_+),\nonumber\\
&& y_\mu =({\bf y}_\perp, y_-, y_+)\equiv ({{\bf y}}, y_+)
\eea

\subsection{Slated Wilson loop dressed with fields}

The undressed  Wilson loop  in the resummed instanton vacuum is

\bea
\label{W1}
\langle {\bf 1}_\theta\rangle =
\langle {\bf W}(\theta, 0_\perp)\,{\bf W}^\dagger(\theta, b_\perp)\rangle_C\approx e^{-Z_+V_C(\xi_\theta)}\nonumber\\
\eea
with

\be
\label{XITHETA}
\xi_\theta=({\rm cos}^2\theta \,b_3^2+b_\perp^2)^{\frac 12}
\ee
and where $V_C(\xi_\theta)\rightarrow V_C(\xi_x)$ follows by analytical continuation $\theta\rightarrow -i\chi$. The spin dressed Wilson loop
to order $1/m_Q^2$ follows by inserting the corrections (\ref{NR}) on the Wilson lines 

\begin{widetext}
\bea
\label{W2}
&&\langle {\bf 1}_\theta\rangle \,\delta_{12}+\bigg(
+\frac{i}{4m_{Q1}^2}\int_{-\frac 12Z_+}^{+\frac 12Z_+} dz_+\bigg[\sigma_{1\alpha\nu}v_\mu \langle  F_{\mu\nu}(x_1, z_+)iD_{\alpha} (x_1,z_+) {\bf 1}_\theta\rangle + 1\leftrightarrow 2\bigg]\nonumber\\
&&+\frac{1}{4m_{Q1}^2}\int_{-\frac 12Z_+}^{+\frac 12Z_+} dz_+\,\int_{-\frac 12Z_+}^{+\frac 12Z_+}dz_+^\prime \bigg[\langle\sigma_{1\mu\nu}F_{\mu\nu}(x_1, z_+)(iD)^2(x_1, z_+^\prime) {\bf 1}_\theta\rangle + 1\leftrightarrow 2\bigg]\nonumber\\
&&+\frac{1}{8m_{Q1}m_{Q2}}\int_{-\frac 12Z_+}^{+\frac 12Z_+}dz_+\, \int_{-\frac 12Z_+}^{+\frac 12Z_+}dz_+^\prime 
\bigg[\langle\sigma_{1\mu\nu}F_{\mu\nu}(x_1, z_+)(iD)^2(x_2, z_+^\prime) {\bf 1}_\theta\rangle \nonumber\\
&&\qquad\qquad\qquad\qquad\qquad\qquad\qquad\qquad +\langle (iD)^2(x_1, z_+)\sigma_{2\mu\nu}F_{\mu\nu}(x_2, z^\prime_+) 
{\bf 1}_\theta\rangle\nonumber\\
&&\qquad\qquad\qquad\qquad\qquad\qquad\qquad\qquad-\frac 12\langle \sigma_{1\mu\nu}F_{\mu\nu}(x_1, z_+)\sigma_{2\alpha\beta}F_{\alpha\beta}(x_2, z^\prime_+) {\bf 1}_\theta\rangle
\bigg]
\bigg)\delta_{12}
\eea
after droping the terms that vanish on-shell, the terms that vanish by parity after averaging in the presence of the undressed
Wilson loop, and those with no contribution to the spin-dependent potentials.   
In (\ref{W2}) we have labeled the quark masses for a general Wilson loop with unequal masses, and used the short hand notation
$$\delta_{12}=\delta({\bf x}_1-{\bf y}_1)\delta({\bf x}_2-{\bf y}_2)$$
Throughout, the affine integration parameters  $z_+, z_+^\prime$ in (\ref{W2}) are proper times. The conversion to ordinary times
$(z_+, z_+^\prime)\rightarrow (z_+, z_+^\prime)/\gamma_E$ amounts to extra Lorentz contraction factors of $1/\gamma_E=\sqrt{1+\dot{x}_E^2}$ 
in Euclidean signature, that will be added at the end by inspection.

\subsection{Identities}

To simplify (\ref{W2}) we use the identities  with slated Wilson lines (dropping the delta functions)

\be
\label{ID1}
{\bf W}(x_+,y_+){\bf W}(y_+,z_+)=\langle x_+|\frac 1{v\cdot D}|y_+\rangle\langle y_+|\frac 1{v\cdot D}|z_+\rangle
=\langle x_+|\frac 1{v\cdot D}|z_+\rangle={\bf W}(x_+, z_+)
\ee
 which is a property of the eikonalized and ordered  Wilson line. More impotantly, we have the identity

\bea
\label{ID2}
&&D_{ \nu}(x_+){\bf W}(x_+,y_+)-{\bf W}(x_+,y_+)D_{ \nu}(y_+)\nonumber\\
&&=\langle x_+|D_{ \nu}\frac 1{v\cdot D}-\frac 1{v\cdot D}D_{ \nu}|y_+\rangle=
\langle x_+|\frac 1{v\cdot D} [v\cdot D, D_{\nu}]\frac 1{v\cdot D}|y_+\rangle\nonumber\\
&&=\langle x_+|\frac 1{v\cdot D} (-iv_\mu F_{\mu \nu})\frac 1{v\cdot D}|y_+\rangle=
\int_{-\frac 12Z_+}^{+\frac 12Z_+} dz_+\,{\bf W}(x_+,z_+)(-iv_\mu F_{\mu\nu})(z_+){\bf W}(z_+,y_+)
\eea
The end-point derivative of  a Wilson line, amounts to an insertion of a pertinent field strength (plaquette in a lattice form) along the line,

\bea
\label{PLAQUETTE}
v_\mu F_{\mu \nu}=v_4F_{4\nu}+v_3 F_{3\nu}={\rm cos}\theta F_{4\nu} +{\rm sin}\theta F_{3\nu}
\eea
Finally, we have the large $|z_+|\rightarrow \infty$ identity 

\be
\label{ID3}
\,{\bf W}(y, z_+; x, z_+ )D_\alpha(x, z_+)\,{\bf W}(x, z_+; y, z_+)\rightarrow \partial_\alpha^y
\ee
as the fields are assumed to vanish at asymptotic $z_+$. 
A repeated use of (\ref{ID1}-\ref{ID3}) allows to simplify (\ref{W2}).



\subsection{First contribution in Eq. \ref{W2}}

Consider the  first contribution in (\ref{W2}) 
without $1\leftrightarrow2$, 

\be
\int_{-\frac 12Z_+}^{+\frac 12Z_+} dz_+\,\sigma_{1\alpha\nu}v_\mu \langle  F_{\mu\nu}(x_1, z_+)iD_{ \alpha} (x_1,z_+) {\bf 1}_\theta\rangle=
\int_{-\frac 12Z_+}^{+\frac 12Z_+} dz_+\,\sigma_{1\alpha\nu}v_\mu \langle  F_{\mu\nu}(x_1, z_+){\bf 1}_\theta\rangle i\partial_{1\alpha}
\ee
\end{widetext}
after using (\ref{ID2}) forward and dropping a vanishing contribution by symmetry. Using again (\ref{ID2}) backward we get

\be
-\sigma_{1\alpha\nu}\partial_{1\nu}\langle {\bf 1}_\theta\rangle \partial_{ 1\alpha}\rightarrow
-\epsilon_{ijk}\sigma_{1k}\partial_{1i}
\langle {\bf 1}_\theta\rangle \partial_{1j}
\ee
Recall that $\sigma_{4i}$ is off-diagonal in the Dirac representation. It drops out on a straight Wilson line, 
with  no particle-anti-particle mixing.  Hence the result

\bea
\label{W3}
\frac{Z_+}{\gamma_E}e^{-Z_+V_C(\xi_0)}\,\epsilon_{ijk}\sigma_{1k}(\partial_{1i}V_C(\xi_\theta))\partial_{1j}
\eea
with the additional Lorentz contraction factor in Euclidean signature, restored.

The analytical continuation of (\ref{W3}) follows by taking $\theta \rightarrow -i\chi$, $\gamma_E\rightarrow \gamma\rightarrow \infty$, with 

\bea
\xi_\theta\rightarrow \xi_x=\sqrt{(\gamma b_3)^2+b_\perp^2}\rightarrow \sqrt{(id/dx/M)^2+b_\perp^2}\nonumber\\
\eea
Hence, 
 
\bea
\label{XX}
\frac 1\gamma \partial_{13} V_C(\xi_x)&=&\frac{\partial V_C(\xi_x)}{\partial{\gamma b_{13}}}\rightarrow \frac{(id/dx_1)}{M\xi_x}V_C^\prime(\xi_x)\nonumber\\
\frac 1\gamma \partial_{13}&=& \frac{\partial}{\gamma\partial x_{13}}\rightarrow \frac {ip_{13}}{\gamma}=is_1m_{Q1}
\eea
are the dominant contributions in (\ref{W3}) at large $\gamma$. The contribution to the squared mass operator is

\begin{widetext}
\bea
\label{SPINORBIT11}
M^2_{LS,C}=2M\bigg[ \frac{\sigma_1\cdot ( {b}_{12}\times s_1\hat 3)}{4m_{Q1}}
-\frac{\sigma_2\cdot ( {b}_{21}\times s_2\hat 3)}{4m_{Q2}} \bigg]\,\frac 1{\xi_x}V_C^\prime(\xi_x)
\eea
\end{widetext}
after symmetrization, and dropping the higher order $1/Mm_Q^2$ contribution.
Here $$b_{21}=(b_2-b_1)_\perp\equiv b_\perp, $$ $s_1={\rm sgn} p_{13}$ is the signum  of the 3-momentum of particle 1
(sign of the helicity), 
and  $x_1$ refers to Bjorken-x for particle 1 ($x$ for particle and $\bar x$ for anti-particle).
This is the light front form of the  spin-orbit potential familiar from atomic physics.

\subsection{Last  contribution in Eq. \ref{W2}}

Consider the spin-spin interaction in (\ref{W2})

\begin{widetext}
\bea
\label{SS1}
-\frac{1}{16m_{Q1}m_{Q2}}\int_{-\frac 12Z_+}^{+\frac 12Z_+}dz_+\, \int_{-\frac 12Z_+}^{+\frac 12Z_+}dz_+^\prime 
\bigg[
\langle \sigma_{1\mu\nu}F_{\mu\nu}(x_1, z_+)\sigma_{2\alpha\beta}F_{\alpha\beta}(x_2, z^\prime_+) {\bf 1}_\theta\rangle
\bigg]
\eea
Since  $\sigma_{4i}$ drops out of the straight Wilson line, the chief contribution in (\ref{SS1}) is

\bea
\label{SS1X}
-\frac{1}{4m_{Q1}m_{Q2}}\int_{-\frac 12Z_+}^{+\frac 12Z_+}dz_+\, \int_{-\frac 12Z_+}^{+\frac 12Z_+}dz_+^\prime 
\sigma_{1i}\sigma_{2j}
\langle B_i(x_1, z_+)B_j(x_2, z^\prime_+) {\bf 1}_\theta\rangle
\eea
The magnetic correlation function in the presence of the Wilson loop ${\bf 1}_\theta$ can be rewritten as follows

\bea
\sigma_{1 i}\sigma_{2 j}\langle B_i(x_1, z_+)B_j(x_2, z^\prime_+) {\bf 1}_0\rangle
=\sigma_{1\perp i}\sigma_{2\perp j}\langle B_{\perp i}(x_1, z_+)B_{\perp j}(x_2, z^\prime_+) {\bf 1}_\theta\rangle
+{\cal O}\bigg(\frac 1{\rm cos\theta}\bigg)\nonumber\\
\eea
\end{widetext}
Here $\vec\sigma_{1,2}=(\sigma_\perp, \sigma_3)_{1,2}$ with particle sub-labeling $1,2$,
and   the notation $\perp=1,2$ (not to be confused with the projection orthogonal to $v_\mu$ above).
The longitudinal contribution of the magnetic field $B_3$ ties to $B_3=-B_-/{\rm cos \theta}$,
 with $B_-=v_\perp\cdot B$ the component orthogonal to ${\bf 1_\theta}$. After analytical continuation
$\theta\rightarrow -i\chi$ the contributions in $1/{\rm cos}\theta$ in (\ref{SS1}) are suppressed by 
${\rm cos} \theta\rightarrow \gamma_E\rightarrow\infty$ and will be dropped.
This is expected since in the infinite momentum frame the transverse components of the gauge fields
$E_\perp, B_\perp$
dwarf the longitudinal ones $E_3,B_3$. 

With this in mind, the result 
for the  2-dimensional and transverse spin-spin potential  prior to the analytical continuation is

\begin{widetext}

\bea
\label{SS3}
&&\frac{\sigma_{1\perp i}\sigma_{2\perp  j}}{4m_{Q1}m_{Q2}}\bigg[\bigg(\hat b_{\perp i}\hat b_{\perp j}-\frac 12\delta_{\perp ij}\bigg)\mathbb V_3(\xi_\theta, \theta)+\frac 12 \delta_{\perp ij}\mathbb V_4(\xi_\theta, \theta)\bigg]
\nonumber\\&&=\frac{\sigma_{1\perp i}\sigma_{2\perp  j}}{4m_{Q1}m_{Q2}}\bigg[\lim_{Z_+ \to \infty}\frac {1}{Z_+\langle {\bf 1}_\theta\rangle \gamma_E^2}
\int_{-\frac 12Z_+}^{+\frac 12Z_+}dz_+\, \int_{-\frac 12Z_+}^{+\frac 12Z_+}dz_+^\prime 
\langle B_{\perp i}(x_1, z_+)B_{\perp j}(x_2, z^\prime_+) {\bf 1}_\theta\rangle\bigg]
\eea
with the Lorentz contraction factor in Euclidean signature $\gamma_E$ restored.
Note the overall sign change in passing from the interaction (\ref{SS1X}) to the potentials (\ref{SS3}).
The analytical continuation in (\ref{SS3}) will be carried explicitly below in the instanton vacuum. 
For general gauge fields, a numerical procedure needs to be developed.

\subsection{Remaining   contributions in Eq. (\ref{W2})}

The remaining contributions in (\ref{W2}) are spin-orbit like. They can be simplified through a repeated use of
the identities (\ref{ID1}-\ref{ID3}),  and the observation that the longitudinal contributions of the gauge fields
$B_3=-B_-/{\rm cos\theta}$ and similarly $E_3=-E_-/{\rm cos \theta}$ drop out after the analytical continuation 
and can be ignored.

The two cross spin-orbit contributions in the last line in (\ref{W2})

\bea
\label{C1}
&&+\frac{1}{8m_{Q1}m_{Q2}}\bigg[\int_{-\frac 12Z_+}^{+\frac 12Z_+}dz_+\, \int_{-\frac 12Z_+}^{+\frac 12Z_+}dz_+^\prime 
\bigg(\langle\sigma_{1\mu\nu}F_{\mu\nu}(x_1, z_+)(iD)^2(x_2, z_+^\prime) {\bf 1}_\theta\rangle \nonumber\\
&&\qquad\qquad\qquad\qquad\qquad\qquad\qquad\qquad +\langle (iD)^2(x_1, z_+)\sigma_{2\mu\nu}F_{\mu\nu}(x_2, z^\prime_+) 
{\bf 1}_\theta\rangle\bigg)\bigg]
\eea
can be simplified. First, we recall that $\sigma_{4i}$ mixes particles and holes and does not contribute to the straight Wilson
world-lines under consideration, so that the relevant contribution in (\ref{C1}) is

\bea
\label{C2}
&&+\frac{1}{4m_{Q1}m_{Q2}}\bigg[\int_{-\frac 12Z_+}^{+\frac 12Z_+}dz_+\, \int_{-\frac 12Z_+}^{+\frac 12Z_+}dz_+^\prime 
\bigg(\langle\sigma_{1k}B_k(x_1, z_+)(iD)^2(x_2, z_+^\prime) {\bf 1}_\theta\rangle \nonumber\\
&&\qquad\qquad\qquad\qquad\qquad\qquad\qquad\qquad +\langle (iD)^2(x_1, z_+)\sigma_{2k}B_k(x_2, z^\prime_+) 
{\bf 1}_\theta\rangle\bigg)\bigg]
\eea
Using the identities (\ref{ID2})  and (\ref{ID3}) we can rearrange the 12-integral in (\ref{C2})

\bea
\label{C3}
&&\int_{-\frac 12Z_+}^{+\frac 12Z_+}dz_+\, \int_{-\frac 12Z_+}^{+\frac 12Z_+}dz_+^\prime 
\langle\sigma_{1k}B_k(x_1, z_+)(iD)^2(x_2, z_+^\prime) {\bf 1}_\theta\rangle\approx\nonumber\\
&&2\sigma_{1k}
\int_{-\frac 12Z_+}^{+\frac 12Z_+}dz_+\, \int_{-\frac 12Z_+}^{+\frac 12Z_+}dz_+^\prime 
\langle B_k(x_1, z_+)z_+^\prime v_\mu F_{\mu j}(x_2, z_+^\prime) {\bf 1}_\theta\rangle i\partial_{2j}
\eea
where only the spin contributing terms are retained.  In deriving (\ref{C3}) we used (\ref{ID3}) to trade $iD$ with $i\partial_2$ at the edge
of the Wilson line, followed by an integration by parts along $z^\prime_+$ using $vD$ and then $[vD, iD]=vF$.
With the analytical continuation in mind, the dominant contribution to the potential stems from $j=3$ and 
$v_4/\gamma_E={\rm cos}\theta/\gamma_E\rightarrow 1$, hence

\bea
\label{C4}
\lim_{Z_+ \to \infty}\frac {1}{Z_+\langle {\bf 1}_\theta\rangle \gamma_E}
\int_{-\frac 12Z_+}^{+\frac 12Z_+}dz_+\, \int_{-\frac 12Z_+}^{+\frac 12Z_+}dz_+^\prime 
\langle B_k(x_1, z_+)z_+^\prime v_4 F_{43}(x_2, z_+^\prime) {\bf 1}_\theta\rangle
\rightarrow \epsilon_{k3i}{b}_{21i}\,\frac 1{\xi_x}{\mathbb V}_2^\prime(\xi_x)
\nonumber\\
\eea
for the interaction, with $b_{21}=-b_{12}\equiv b_\perp$.
If we recall the sign flip in passing from the interaction vertex to the potential,
the  $12+21$ spin-orbit contribution to the squared mass operator is

\bea
\label{C5}
M^2_{LS,12}&=&2M\bigg(
-\frac{1}{4m_{Q1}m_{Q2}}\bigg[2\sigma_{1k}\epsilon_{k3i}{b}_{21i} \frac{i\partial_{23}}{\gamma}+1\leftrightarrow 2\bigg]
\frac 1{\xi_x}\mathbb V_2^\prime(\xi_x)\bigg)\nonumber\\
&=&2M\bigg(
\bigg[\frac{\sigma_2\cdot(b_{12}\times s_1\hat 3)}{2 m_{Q2}}-\frac{\sigma_1\cdot(b_{21}\times s_2\hat 3)}{2 m_{Q1}}\bigg]\frac 1{\xi_x}
\mathbb V_2^\prime(\xi_x)\bigg)
\eea

Using similar arguments, the spin-orbit contribution in the second line of (\ref{W2})  yields the dominant contribution 
to the interaction 

\bea
\label{C4X}
\lim_{Z_+ \to \infty}\frac {1}{Z_+\langle {\bf 1}_\theta\rangle \gamma_E}
\int_{-\frac 12Z_+}^{+\frac 12Z_+}dz_+\, \int_{-\frac 12Z_+}^{+\frac 12Z_+}dz_+^\prime 
\langle B_k(x_1, z_+)(z_+^\prime -z_+) v_4 F_{43}(x_1, z_+^\prime) {\bf 1}_\theta\rangle
\rightarrow \epsilon_{k3i}{b}_{21i}\,\frac 1{\xi_x}{\mathbb V}_1^\prime(\xi_x)
\nonumber\\
\eea
with the corresponding $11+22$ spin-orbit contribution to the squared mass operator

\bea
\label{C5X}
M^2_{LS,11}&=&2M\bigg(
-\bigg[\frac{2}{4m^2_{Q1}}\sigma_{1k}\epsilon_{k3i}{b}_{21i} \frac{i\partial_{13}}{\gamma}+1\leftrightarrow 2\bigg]
\frac 1{\xi_x}\mathbb V_1^\prime(\xi_x)\bigg)\nonumber\\
&=&2M\bigg(
\bigg[\frac{\sigma_1\cdot (b_{12}\times s_1\hat 3)}{2 m_{Q1}}-\frac{\sigma_2\cdot (b_{21}\times s_2\hat 3)}{2 m_{Q2}}\bigg]
\frac 1{\xi_x}\mathbb V_1^\prime(\xi_x)\bigg)
\eea
\end{widetext}

Below, we explicitly show how to evaluate $\mathbb V_{1,2}(\xi_x)$ in the instanton vacuum. For general gauge fields,
a numerical procedure needs to be developed, as we noted earlier for the spin-spin interaction (\ref{SS3}).

\section{Light front hamiltonian in the instanton vacuum}
\label{SPIN-NZMF}

For the particular case of the instanton vacuum, these spin potentials are essentially generated by non-zero modes
(those due to the zero modes will be discussed below). They are related to the central 
electric potential $V_{C}(\xi_x)$ in (\ref{16Z}),  since  the induced spin correlators satisfy $BB=EE$ and $BE=\pm EE$ by self-duality.

\subsection{Spin-Spin interaction}

More specifically, the spin-spin interaction (\ref{SS3})  with self-dual fields,  reads

\begin{widetext}
\bea
\label{SS3X}
-\frac{\sigma_{1\perp i}\sigma_{2\perp j}}{4m_{Q1}m_{Q2}}\bigg[\lim_{Z_+ \to \infty}\frac 1{Z_+\langle {\bf 1}_\theta\rangle\gamma_E^2}
\int_{-\frac 12Z_+}^{+\frac 12Z_+}dz_+\, \int_{-\frac 12Z_+}^{+\frac 12Z_+}dz_+^\prime 
\langle E_{\perp i}(x_1, z_+)E_{\perp j}(x_2, z^\prime_+) {\bf 1}_\theta\rangle\bigg]
\eea
\end{widetext}
with here $i,j=1,2$. With this in mind, we now note that (\ref{PLAQUETTE}) amounts to

\bea
\label{FTOE}
v_\mu F_{\mu i}=-R_{ij} (\theta) E_j
\eea
using again the self-duality for the instanton, with the rotation matrix

\bea
R(\theta)=
\begin{pmatrix}
{\rm cos}\theta & -{\rm sin}\theta\\
{\rm sin}\theta & {\rm cos}\theta
\end{pmatrix}\rightarrow 
\gamma R=\gamma
\begin{pmatrix}
1& +i\\
-i& 1
\end{pmatrix}\nonumber\\
\eea
and its analytical continuation. Note that for the anti-instanton which is anti-self-dual,  $R(\theta)\rightarrow R(-\theta)$
and $R\rightarrow R^*$.
Inserting the inversion of (\ref{FTOE}) in (\ref{SS3X}) gives

\begin{widetext}
\bea
\label{SS3XX}
-\frac{\sigma_{1\perp i}\sigma_{2\perp j}}{4m_{Q1}m_{Q2}}\bigg[\lim_{Z_+ \to \infty}\frac {R_{im}(\theta)R_{jn}(\theta)}{Z_+\langle {\bf 1}_\theta\rangle \gamma_E^2}
\int_{-\frac 12Z_+}^{+\frac 12Z_+}dz_+\, \int_{-\frac 12Z_+}^{+\frac 12Z_+}dz_+^\prime 
\langle v_\mu F_{\mu m}(x_1, z_+)v_\nu F_{\nu n}(x_2, z^\prime_+) {\bf 1}_\theta\rangle\bigg]\nonumber\\
\eea
Using twice the identity (\ref{ID2}) allows to simplify (\ref{SS3XX})

\bea
\label{SS3XXX}
-\frac{\sigma_{1\perp i}\sigma_{2\perp j}}{4m_{Q1}m_{Q2}}\bigg[\lim_{Z_+ \to \infty}\frac {-R_{im}(\theta)R_{jn}(\theta)\partial^1_{ m}\partial^2_{ n}\langle {\bf 1}_\theta\rangle}{Z_+\langle {\bf 1}_\theta\rangle\gamma_E^2}\bigg]
\eea
Recalling the overall sign flip in passing from the interaction to the potentials,  the analytical continuation of (\ref{SS3XXX}) gives the 
instanton spin-spin contribution to the squared mass operator

\bea
\label{SS3XXXX}
M^2_{SS}&=&2M\bigg(
 \frac{\sigma_{1\perp i}\sigma_{2\perp j}}{4m_{Q1}m_{Q2}}\,\bigg[\bigg(\hat b_{12i}\hat b_{12 j}-\frac 12\delta_{\perp ij}\bigg)\mathbb V_3(\xi_x)+\frac 12 \delta_{\perp ij}\mathbb V_4(\xi_x)\bigg]\bigg)\nonumber\\
&=&2M
\bigg( \frac{\sigma_{1\perp i}\sigma_{2\perp j}}{4m_{Q1}m_{Q2}}\,\bigg[R_{im}R_{jn}\partial_{1m}\partial_{1n} V_C(\xi_x)\bigg]\bigg)
\eea
\end{widetext}
Hence the relation of $\mathbb V_{3,4}(\xi_x)$ to the central Coulomb potential $\mathbb V_C(\xi_x)$ induced by  instantons
(anti-instantons) 

\bea
\label{V12X}
\mathbb V_3(\xi_x)&=&\frac{2b_\perp^2}{\xi_x^2}\mathbb V_C^{\prime\prime}(\xi_x)\nonumber\\
\mathbb V_4 (\xi_x)&=&0
\eea

In the instanton vacuum, the  light front spin-orbit potentials $\mathbb V_{1,2}$ in   (\ref{C4}-\ref{C4X}), can be shown to be tied by the same
identity as their counterparts in the rest frame~\cite{Eichten:1980mw}, namely

\bea
\label{V12C}
\mathbb V_2(\xi_x)=\mathbb V_1(\xi_x)+V_C(\xi_x) \rightarrow \frac 12 V_C(\xi_x)
\eea
with the rightmost result following in the instanton vacuum.
Indeed, while on the light front $\mathbb V_{2,4}(\xi_x)$  are no longer tied by the Bianchi-identity (covariantized Lenz law),
we note that  the leading contributions in (\ref{C4}-\ref{C4X}) match the rest frame contributions at $\theta=0$. Therefore, the rest frame
relation $\mathbb V_2(R)=\frac 12 V_C(R)$ in the instanton vacuum~\cite{Eichten:1980mw} (note the sign convention difference), carries to the light front $\mathbb V_2(\xi_x)=\frac 12 V_C(\xi_x)$. This is not the case for $\mathbb V_{3,4}(\xi_x)$ as we have shown.

\subsection{Light front hamiltonian}

The sum of the spin contributions  to the squared mass operator on the light front in the 
instanton vacuum, is now explicit and of the form

\begin{widetext}
\bea
\label{LFHAMIL}
M^2_{SD,I}(\xi_x, b_\perp)=2MV_{SD,I}(\xi_x, b_\perp)=2M&&
\bigg(
\bigg[ \frac{\sigma_1\cdot ( {b}_{12}\times s_1\hat 3)}{4m_{Q1}}
-\frac{\sigma_2\cdot ( {b}_{21}\times s_2\hat 3)}{4m_{Q2}} \bigg]\,\frac 1{\xi_x}V_C^\prime(\xi_x)\nonumber\\
 &&+
\bigg[\frac{\sigma_1\cdot (b_{12}\times s_1\hat 3)}{2 m_{Q1}}-\frac{\sigma_2\cdot (b_{21}\times s_2\hat 3)}{2 m_{Q2}}\bigg]
\frac 1{\xi_x}\mathbb V_1^\prime(\xi_x)\nonumber\\
&&+
\bigg[\frac{\sigma_2\cdot(b_{12}\times s_1\hat 3)}{2 m_{Q2}}-\frac{\sigma_1\cdot(b_{21}\times s_2\hat 3)}{2 m_{Q1}}\bigg]\frac 1{\xi_x}
\mathbb V_2^\prime(\xi_x)\nonumber\\
&&+
\frac 1{4m_{Q1}m_{Q2}}\sigma_{1\perp i}\sigma_{2\perp j}\bigg[\bigg(\hat b_{21i}\hat b_{21j}-\frac 12\delta_{\perp ij}\bigg)\mathbb V_3(\xi_x)
\bigg]\bigg)\nonumber\\
\eea
\end{widetext}
with $b_{21}=-b_{12}=b_\perp$  and $s_{1,2}$ the signum of the velocity along the 3-direction (sign of the helicity).
The contributions in (\ref{LFHAMIL}) are in (\ref{SPINORBIT11}), (\ref{C5}) and (\ref{C5X}).
All spin potentials $\mathbb V_{1,2,3}(\xi_x)$ are tied to the central potential $\mathbb V_C(\xi_x)$ in the instanton vacuum,
as   in (\ref{V12X}) and (\ref{V12C}).  

A key feature of the spin orbit contributions in (\ref{LFHAMIL}), is that a flip of a spin say $\sigma_1$,
can be compensated by a flip in the sign of the helicity say $s_1$ or $s_2$. This is reminiscent of the rest frame symmetry of the spin-orbit interactions,
that show that a flip in the spin can be compensated by a flip in the angular momentum, thereby preserving the total angular momentum.

The  light front hamiltonian in the instanton vacuum,  is the squared mass operator for a $Q\bar Q\equiv Q_1Q_2$ pair, that includes the free  plus the central contribution 
in (\ref{16Z}),  and the spin contributions (\ref{LFHAMIL}), 

\bea
\label{MASS2}
M^2=&&\sum_{i=1,2}\frac{k_\perp^2+m^2_{Qi}}{x_i}\nonumber\\
&&+2M (V_{C}(\xi_x)+V_{SD,I}(\xi_x, b_\perp))\nonumber\\
\eea
with Bjorken $x_{i=1,2}$ and satisfying  $x_1+x_2=1$. A detailed analyses of the spectrum and light front wavefunctions
following from (\ref{MASS2}) as applied to heavy and light mesons,
with comparative estimates from perturbative one-gluon exchange and confinement, will be detailed in a sequel.

\section{Spin interaction from a string on the light front}
\label{I_spin_orbit}


In the rest  frame, the spin-dependent contributions emerging from the string were
discussed by Buchmuller~\cite{Buchmuller:1981fr} and others~\cite{Pisarski:1987jb,Gromes:1984ma}. Since the spin-spin interactions are short ranged, 
only the {\it self} spin-orbit contributions survive at large separation $R$, where the string is active. Also,
the electric flux tube is confined to the string, so the {\it self} spin-orbit contribution is
mostly induced by Thomas precession which is of {\it opposite sign} to the spin-orbit contribution from the standard Coulomb field. 

More specifically, in the rest frame and at large separation
only the {\it self} spin-orbit potential survives~\cite{Buchmuller:1981fr,Pisarski:1987jb,Gromes:1984ma}

\begin{widetext}
\bea
\label{STRING1X}
V_{LS,{\rm string}}(R)&\approx&\bigg(\frac{\sigma_1\cdot L_1}{4m_{Q1}^2}-\frac{\sigma_{2}\cdot L_{2}}{4m_{Q2}^2}\bigg)
\bigg(\frac 1R V^\prime_C(R)+\frac 2R V^\prime_1(R)\bigg)\nonumber\\
&\approx&
\bigg(\frac{\sigma_1\cdot L_1}{4m_{Q1}^2}-\frac{\sigma_{2}\cdot L_{2}}{4m_{Q2}^2}\bigg)
(1 -2) \frac {\sigma_T}R 
\eea
with the convention for the orbital angular momenta $L_1=-L_2=L$. Here, 
$V_C(R)=\sigma_T R$, and $V_1(R)\approx -V_C(R)$ from  (\ref{V12C}),  since the {\it cross} spin-orbit potential 
being short ranged, is expected to vanish at large $R$, i.e. $V_2(R)\approx 0$. 
On the light front, (\ref{STRING1X}) can be recast, and its contribution to the squared mass operator is

\bea
\label{SLSTRING}
M^2_{LS,\rm string}\approx 2M\bigg(
\bigg[\frac{\sigma_1\cdot({b}_{12}\times s_1\hat 3)}{4 m_{Q1}}
-\frac{\sigma_2\cdot({b}_{21}\times s_2\hat 3)}{4 m_{Q2}}
\bigg]
(1-2)\frac{\sigma_T}{\xi_x}\bigg)
\eea
\end{widetext}
after using (\ref{V12C}), and borrowing from the  spin reduction  structure in  (\ref{SPINORBIT11})  and (\ref{C5X}).
As we noted in~\cite{Shuryak:2021fsu} (see Appendix B), the spin-orbit potential following from the analysis in~\cite{Eichten:1980mw} which we have followed
(for both the instantons and string), is twice larger~\cite{Pineda:2000sz}.

Finally, we note that on the light front, the sign of the string induced {\it self} spin-orbit in (\ref{SLSTRING}) is 
similar to the one expected from instantons in the dense regime,  but opposite to the sign following from the perturbative Coulomb exchange,
as originally noted in the rest frame in~\cite{Buchmuller:1981fr,Pisarski:1987jb,Gromes:1984ma}.

\section{Spin-flavor interactions  for light quarks}
\label{t_HOOFT}

The spin-flavor interactions for light quarks are well understood in the rest frame. They involve chiefly the fermionic {\it zero modes} as they
tunnel through instantons and anti-instantons. Because of the Pauli principle, only the zero modes with different flavors can undergo 
simultaneous tunneling, resulting in the famed $^\prime$t Hooft interactions. For three flavors and in the zero  size approximation,
the 3-flavor interaction is repulsive and mostly active in the flavor singlet channel~\cite{Chernyshev:1995gj,Shuryak:2021iqu}

\begin{widetext}
\begin{align}
\label{21X}
{\cal V}^{L+R}_{qqq}=&\frac {G_{Hooft}}{N_c(N_c^2-1)}
\bigg(\frac{2N_c+1}{2(N_c+2)}\,{\rm det}(UDS)\nonumber\\
&+\frac 1{8(N_c+1)}\,
\bigg({\rm det}(U_{\mu\nu}D_{\mu\nu}S)+{\rm det}(U_{\mu\nu}DS_{\mu\nu})+{\rm det}(UD_{\mu\nu}S_{\mu\nu})\bigg)\bigg) +(L\leftrightarrow R)
\end{align}
\end{widetext}
with a strength

\be
\label{22X}
G_{Hooft}=\frac{{n_{I+\bar I}}}2\bigg(\frac{4\pi^2\rho^3}{m_Q\rho}\bigg)^3
\ee
and the short hand notations   ($Q\equiv U,D,S$)

\be
\label{23X}
Q=\overline q_Rq_L,\,\,\, Q_{\mu\nu} =\overline q_R\sigma_{\mu\nu}q_L,\,\,\, Q^a=\overline q_R\sigma^aq_L
\ee
The 2-flavor ud-interaction is attractive and follows by vacuum averaging the s-contribution. It is also determinantal

\bea
\label{24X}
{\cal V}^{L+R}_{qq}=&&\,{\kappa}_2\,A_{2N}\,
\bigg({\rm det}(UD)+B_{2N}\,{\rm det}(U_{\mu\nu}D_{\mu\nu})\bigg)\nonumber\\
&&+(L\leftrightarrow R)
\eea
and  attractive $$\kappa_2=3! G_{Hooft}\langle \overline s_R s_L\rangle=3 G_{Hooft} \langle \overline s s\rangle<0,$$ 
\bea
\label{25X}
A_{2N}=\frac{(2N_c-1)}{2N_c(N_c^2-1)}\qquad
B_{2N}=\frac 1{4(2N_c-1)}\nonumber\\
\eea
In the Weyl basis 
$\sigma_{\mu\nu}\rightarrow  i\eta^a_{\mu\nu}\sigma^a$ with the $^\prime$t Hooft symbol satisfying $\eta^a_{\mu\nu}\eta^b_{\mu\nu}=4\delta^{ab}$, (\ref{21X})
can be simplified 
\bea
\label{26X}
{\cal V}^{L+R}_{qq}=&&\,{\kappa}_2\,A_{2N}\,
\bigg({\rm det}(UD)-4B_{2N}\,{\rm det}(U^aD^a)\bigg)\nonumber\\
&&+(L\leftrightarrow R)
\eea
In the rest frame, (\ref{26X}) contributes an  ultra-local interaction potential.
In leading order in $1/N_c$, the potentail in the U(1) or $\eta^\prime$ channel, is

\bea
\label{XX1}
\frac 12 \kappa_2 A_{2N}\frac 12 \bigg({1-\tau_1\cdot \tau_2}\bigg)\,\delta(\vec x_{12}) 
\eea
 with  $\vec x_{12}=\vec x_1-\vec x_2$. The corrections to (\ref{XX1}) are non-relativistic  $\nabla/m_Q$.

The corresponding pair interaction  on the light front in the eikonalized approximation,
can be written schematically as

\begin{widetext}
\bea
\label{27X}
{\cal V}_{12}^{L+R}(\tilde\xi_x)\rightarrow \frac 12 \tilde\kappa_2 A_{2N}\frac 12 \bigg({1-\tau_1\cdot \tau_2}\bigg)\,
\sigma_{\perp 1}\cdot\sigma_{\perp 2}\,\delta(\tilde\xi_x) 
\equiv {\cal V}^{L+R}_{\eta^\prime}\delta(\tilde\xi_x)
\eea
\end{widetext}
in the $\eta^\prime$ channel, with $\tilde\kappa_2=\kappa_2/\rho^3$. We dropped the spin-independent mass contributions.
in writing (\ref{27X}) the delta function is assumed to depend  only on the invariant 1D-like distance $\tilde \xi_x$ defined in (\ref{W99X}),
in the local approximation. The alternative delta function  $$\delta(\tilde\xi_x)\rightarrow \delta (P^+z^-)\delta(x_\perp/\rho)$$
which is 3D-like and local,  will be discussed elsewhere.
The  flavor permutation inherent to the flavor singlet $^\prime$t Hooft vertex  is manifest in

\be
\label{28X}
1-{\bf P}_{12}=\frac 12(1-\tau_1\cdot \tau_2)
\ee
with ${\bf P}_{12}$ the flavor pair permutation operator. 
The spin-flip interaction  $\sigma_{\perp 1}\cdot\sigma_{\perp 2}$ remains on the light front in the near mass-shell limit,
and  flips the helicity of the incoming quark pair from L-left to R-right in the instanton contribution, and vice-versa in the anti-instanton 
contribution. The corresponding interactions in the
scalar and pseudoscalar  channels  follow by Fierzing. There is no induced interaction by Fierzing 
in  the vector and pseudo-vector channels. In particular, it is 
attractive in the pion channel and zero in the rho channel ($N_c=3$)

\bea
\label{29X}
&&{\cal V}_{\pi}^{L+R}(\tilde\xi_x) \approx \tilde{\kappa}_2\,\delta(\tilde\xi_x)\nonumber\\
&&{\cal V}_{\rho}^{L+R}(\tilde\xi_x) \approx 0
\eea
\begin{widetext}
For light quarks solely in the instanton vacuum, the light  front mass operator (\ref{16Z}) now reads

\be
\label{16} 
M^2\approx  \frac{k_\perp^2+m_Q^2}{{x\bar x}} +2M \bigg[\bigg(\frac{4\kappa }{N_c \rho}\bigg){\bf H}(\tilde\xi_x)+ {\cal V}_{P}^{L+R}(\tilde\xi_x)\bigg]
\ee
\end{widetext}
where $P$ refers to the non-vanishing Fierz contributions ($P=\pi, \sigma, \pi_5, \sigma_5, \eta^\prime$), with the constituent quark mass $m_Q$ added.  
The instanton contributions and the constituent mass are of order $\kappa$ in the packing fraction.

\section{The pion on the light front}
\label{pion}

The spontaneous breaking of chiral $SU(N_f)_A$ symmetry is  a fundamental and  important phenomenon
of nonperturbative QCD,  much like  confinement. The pion plays  a very special role 
in it, being a Nambu-Goldstone (near massless) mode, a wave riding the vacuum quark condensate.
The question then is: 
how does the pion  emerge  from (\ref{16}) on the light front,  with a {\it non-vanishing} constituent quark
mass $m_Q$? 
We will return to the pion wave function, including its quark spin-orbital wave functions, in our next publication.
Here we give a qualitative description of the pion on the light front,  in a vacuum randomly populated with only instantons and anti-instantons.

 Let us start by noting that as $M\rightarrow 0$, then 
$\tilde\xi_x\rightarrow 1/M\rho/|id/dx|\gg 1$,  and  ${\bf H}(\tilde\xi_x)$ is dominated by its large asymptotic constant in  (\ref{17Z}),
which should be removed because of confinement. With this in mind, (\ref{16}) simplifies
\bea
\label{16W0} 
M^2\approx  \frac{M_\perp^2}{x\bar x}+2M^2\,{\cal V}_\pi^{L+R}\,\delta(M\tilde\xi_x) 
\eea
with $M_\perp^2= {k_\perp^2+m_Q^2}$. It is iterative and geometrical (reminiscent
of a Schwinger-Dyson equation).
As a result, the squared mass operator is  formally

\begin{widetext}
\bea
\label{16W4}
M^2\approx \bigg[\frac 1{\sqrt{x\bar x}}\bigg(1-2\rho{\cal V}^{L+R}\delta(|id/dx|)\bigg)^{-1}\frac 1{\sqrt{x\bar x}}\bigg]\,M^2_\perp\nonumber\\
\eea
\end{widetext}
where the  ordering ambiguity is  fixed  by symmetrization,  to enforce hermiticity.  (\ref{16W4}) admits a normalizable massless state
\be
\label{PIONWF}
\varphi_\pi(x, b_\perp)=(6x\bar x)^{\frac 12}\, \bigg(\frac{m_Q}{\sqrt\pi}K_0(m_Q |b_\perp|)\bigg)
\ee
with $\int dx db_\perp \varphi_\pi^2(x, b_\perp)=1$, at a resolution of $1/\rho$.

When a small current mass $m_q$ is added, the constituent mass
shifts in leading order $m_Q\rightarrow m_Q+m_q$, and the pion becomes massive

\bea
\label{16W5}
M_\pi^2= && 2m_qm_Q\int_{0}^{1}dx\frac{\varphi^2_\pi(x)}{x\bar x}+{\cal O}(m_q^2)\nonumber\\
= &&12m_qm_Q+{\cal O}(m_q^2)
\eea
as expected for a Goldstone mode. In the random instanton vacuum (RIV), the constituent mass $m_Q$ follows from the breaking of chiral symmetry with
explicitly
$$\langle\bar q q\rangle=-{N_cm_Q}/{2(\pi\rho)^2}$$ (see for instance Eq. 84 in~\cite{Shuryak:2021fsu}), and (\ref{16W5}) reduces to the GOR relation

\be
\label{16W6}
M_\pi^2= -2m_q\bigg[\frac{\langle \bar q q\rangle}{f_\pi^2}\bigg]+{\cal O}(m_q^2)
\ee
The squared pion decay constant is identified as 
\bea
f_\pi^2=N_c/(\pi\rho)^2/12=1/(2\pi\rho)^2
\eea
 The numerical value is  surprisingly accurate, with 
$f_\pi\sim 96\,{\rm MeV}$ for $\rho\sim \frac 13\,{\rm fm}$.

Note that in our case the pion  -- as a true Goldstone mode--  is massless, eventhough the constituent quark mass $m_Q\neq 0$. The would-be-pion in 1+1 dimension becomes massless only in the large $N_c\rightarrow \infty$ limit.

\begin{figure}[h!]
	\begin{center}
		\includegraphics[width=7cm]{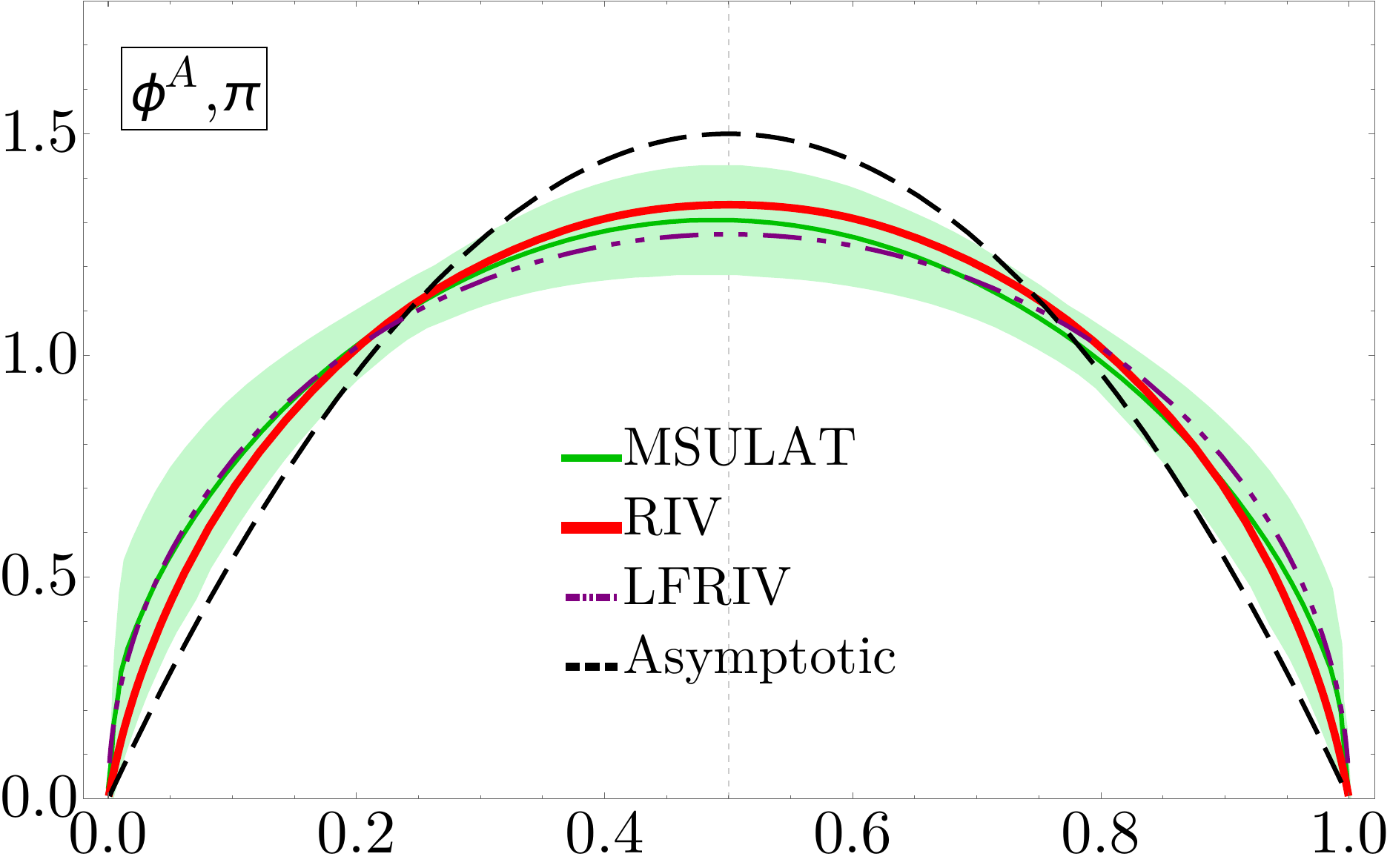}
		\caption{ RIV is the pion longitudinal DA from the  instanton vacuum  in the rest frame~\cite{Kock:2021spt}; LFRIV is the pion DA from the instanton vacuum on the light front;
		MSULAT is the pion DA from the lattice~\cite{Zhang:2020gaj}; the asymptotic pion DA of $6x\bar x$ is shown for comparison.}
		\label{fig_pipda}
	\end{center}
\end{figure}

 In Fig.~\ref{fig_pipda} we show  the  longitudinal pion distribution amplitude (DA) in the random  instanton vacuum on the light front LFRIV versus Bjorken $x$,
following from (\ref{PIONWF}) with  $\varphi^{A}_\pi(x)=\frac 8\pi \sqrt{x\bar x}$ normalized as  $\int dx \varphi^{A}_\pi(x)=1$, at the low resolution of $1/\rho$. It is in  good agreement with the 
pion DA in the random instanton vacuum RIV  obtained in  the rest frame, at the same resolution~\cite{Kock:2021spt}. Both results are compared to the asymptotic pion DA of $6x\bar x$ dashed curve,
 and  the lattice pion DA MSULAT green-curve-band~\cite{Zhang:2020gaj}. LFRIV is similar to the one derived using the Schwinger-Dyson construction
~\cite{Roberts:2021nhw}, and identical to the pion longitudinal DA  discussed using  light-front holography~\cite{DeTeramond:2021jnn}.

The pion DA is driven by
the spontaneous breaking of chiral symmetry, which is the same whether in the rest frame or in the light cone frame,
and as expected is not sensitive to the confinement mechanism. 
What is sensitive to confinement are the pion excited states,
for instance the radial $\pi(1300)$ excitation with assignments $1^-(0^{-+})$ and higher, as we discussed earlier.
In this case, the instanton contribution in (\ref{16}) is only asymptotic and the role of confinement as in (\ref{LIGHT3}) is  important.
The  relevant squared mass operator on the light front  is now

\bea
\label{16W7} 
M^2\approx\frac{M_\perp^2}{{x\bar x}} +2M\big[{\rm spin}\big]+2\sigma_TM\xi_x
\eea
The short range spin interactions due to the instantons non-zero modes are given in~\ref{SPIN-NZMF}.

\section{Conclusions}
\label{conclusions}

We started this paper by discussing the basic problem of two massive  relativistic quarks, connected by
 a  classical Nambu-Gotto string. The problem was
 first set in the CM frame, in which the spectrum and the wave functions are readily obtained,
by solving (semiclassically or directly) a relativistic  Klein-Gordon  equation.

 Then we considered the same problem 
 on the light front (without spin effects),
 and derived the ensuing Hamiltonian. After turning it to a quadratic form (using the einbein trick), 
 we defined an appropriate functional basis in which
 its diagonalization  can be carried explicitly. It yields a meson spectrum that is consistent with the one
from the rest frame, and  with the expected and observed Regge behavior in terms of Regge slope.
The Regge intercept turned out to be higher than the observed one, at least for the vector mesons we discussed.

The LFWFs of the low-lying states are obtained, as a function of the transverse momentum $\rho=p_\perp$ and Bjorken $x$, see e.g. (\ref{LFWF_1}) for the ground state.
To a  good approximation,  they are dominated by the lowest harmonic of the pertinent diagonalization  set,
a Gaussian in the $p_\perp$ direction,  and a $sin(\pi x)$ in the longitudinal direction.

A massless left-handed quark tunneling through an instanton
emerges as a right-handed massless quark as a zero mode, a remarkable feature of a vector interaction. This is the essence of the dynamical
breaking of chiral symmetry, which gives a running constituent mass. The collectivization of these zero modes is well undertstood in the rest frame, and
yields the octet of massless Golstone modes. The QCD vacuum in the Zero Mode Zone is $^\prime metallic^\prime$, with the scalar and 
vector mesons as weakly correlated  $^\prime excitons^\prime$. Their orbitally excitations are sensitive to confinement.

The role of the non-perturbative vacuum structure on the light cone,  is best seen by noting that all hadron correlators on the light cone
map onto a  Wilson loop sloped at an angle $\theta$ in Euclidean space, that analytically continues to $-i\chi$ the rapidity in Minkowski
space, a proposal we made long ago. An excited and confined meson wether light or heavy, is characterized by a staight string
with massive end-points, to account for the scalar masses from the spontaneous breaking of chiral symmetry, plus current masses. In this 
sense all  mesons  behave democratically on the light cone.

The role of instantons and anti-instantons on the mesons in the light cone,  follows from the parallell Wilson lines before analytical continuation.
Their effects fall into two categories: 1/ the non-zero modes and 2/ the zero modes. The contribution of the non-zero modes can be explicitly calculated
using the sloped Wilson loop, and then analytically continued to Minkowski space, giving rise to a central and spin contributions to the mass operator. The
contribution of the zero-modes is still captured by the local form of the $^\prime$t Hooft pair interaction, in addition to the constituent quark mass.
We have explicitly assessed  these contributions, and
derived the pertinent mass operator in Minkowski signature. Modulo ordering ambiguities, it is iterative and non-local.

For a tightly bound pion
where confinement is less active in a vacuum dominated by instantons, 
we have shown that this operator admits  an exact Golstone mode on the light front,  with the correct GOR relation
and pion decay constant, and a  universal and normalizable light-front  DA. The latter is in good agreement with the one derived 
in the rest frame, using the quasi-DA construction.

The role of the spin effects on the light-light, heavy-light and heavy-heavy mesons on the light front 
will be discussed next.

\vskip 1cm
{\bf Acknowledgements}

This work is supported by the Office of Science, U.S. Department of Energy under Contract No. DE-FG-88ER40388.

\appendix
\section{Klein-Gordon equation and the lowest meson states}
\label{KLEIN}

Now we return to the quantized version of the relativistic wave equation (\ref{eqn_dispersion}). Unlike the
nonrelativistic Schroedinger equation, the energy $E$ does not appear in it linearly,  but is solvable. The problem with it, is related with the solution behavior at large distances, beyond the turning  point $r>r_*$. This is a textbook situation for  the Schroedinger equation, where  $p^2$ changes sign and therefore $p$ becomes, imaginary. The correct solution decays exponentially in this region, hence the quantization
condition $\psi(r\rightarrow \infty)=0$.

The Klein-Gordon equation in this situation leads to a change of sign for $\sqrt{p^2}$, leading to complex $p$, so
at large $r$ the solutions are oscillating with increasing frequency. Physically,  this corresponds to
the acceleration of produced quanta in a constant electric field, which obviously has no relation with
the confining string problem we are after. This is at the origin of the Klein-paradox following from
pair creation.

As an  approximation,  one  may choose  to consider the wave function only inside the ``normal region" $r<r_*$,
and use  as a quantization condition the wave function vanishing at the turning point $\psi(r_*)=0$.

\begin{figure}[t]
	\centering
	\includegraphics[width=7cm]{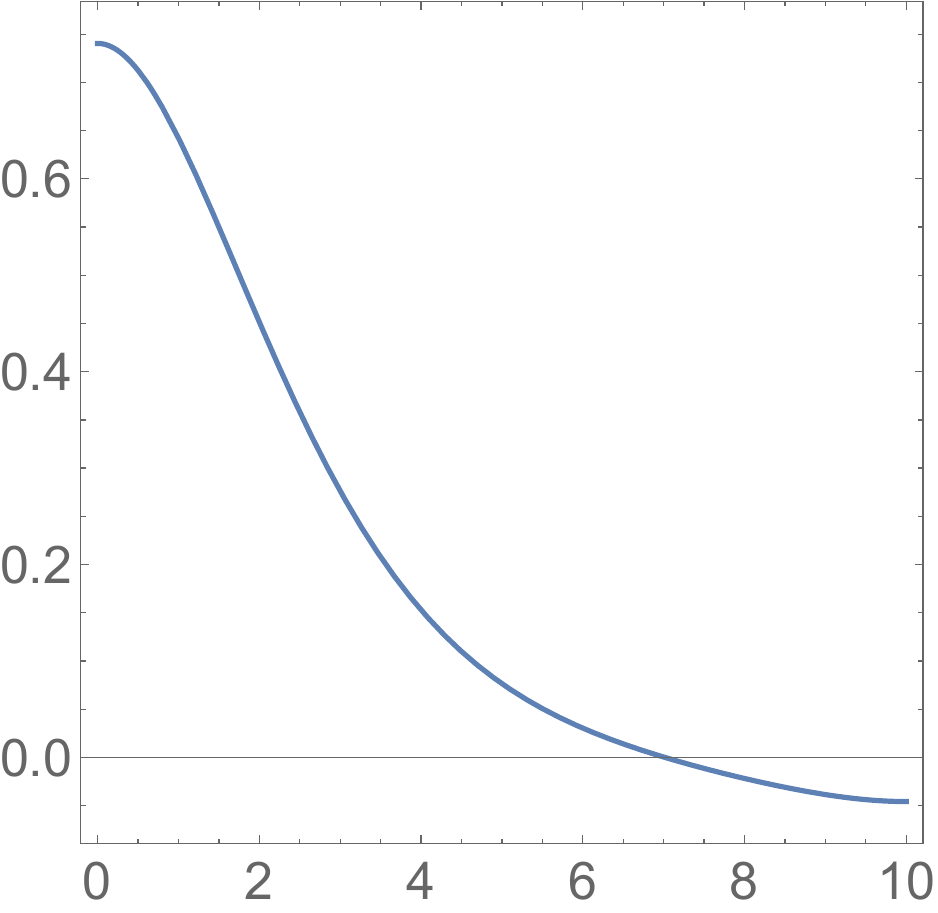}
		\includegraphics[width=7cm]{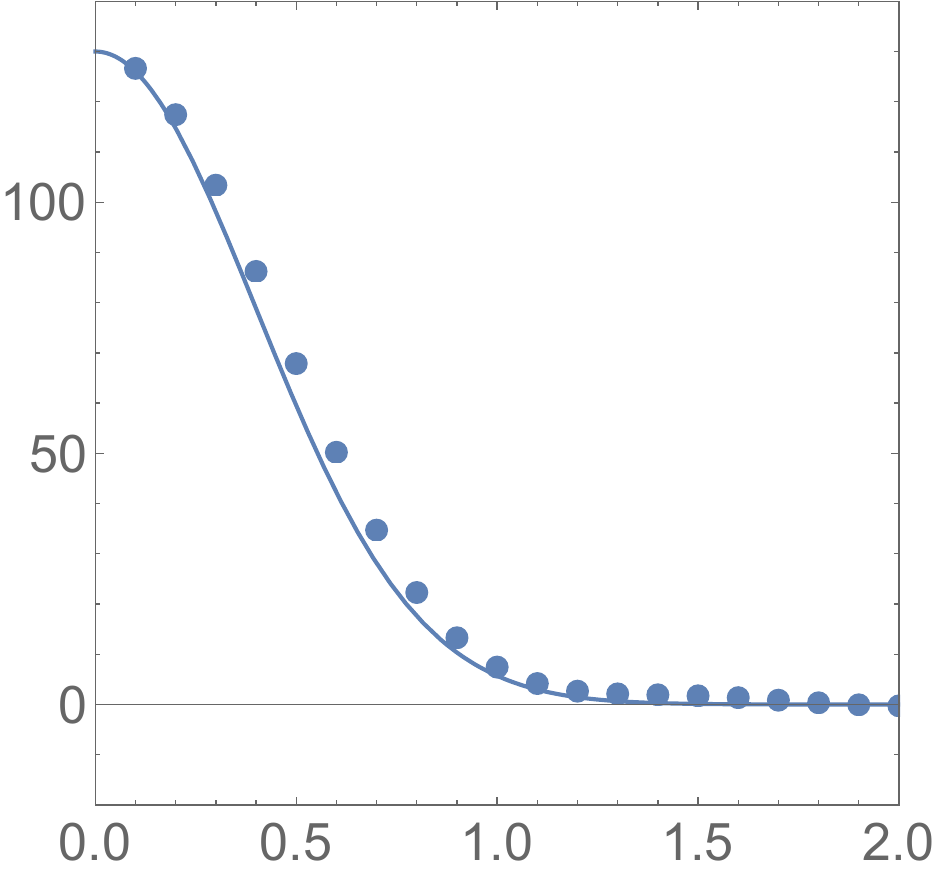}
	\caption{The ground state wave function $\psi_0(r)$ versus $r (GeV^{-1})$ (upper plot)
	and its Fourier transform $\psi_0(p)$ (points, lower plot) versus moomentum $p (GeV)$. The line is a Gaussian  shown for comparison.}
	\label{fig:psi0}
\end{figure}

For the constituent quark mass and string tension, the ground state
wave function is shown in Fig.~\ref{fig:psi0}. The energy in the Klein-Gordon  equation, corresponding to half
or reduced system is 1.06 $GeV$, so the total mass is $M_0=2.12\, GeV$.
Its Fourier transform (also calculated for $r<r_*$)
is shown  in the lower plot for $\psi_0(p)$. For comparison we show a Gaussian fit with
$\langle p^2 \rangle^{1/2}=0.33\, GeV$. 

\section{Basis functions}
\label{basis}

\subsection{The transverse oscillator}

As explained in the text, we use the set of two-dimensional oscillator functions,
with transverse momenta as argument. One may either use a double set of 
standard one-dimensional oscillator functions in cartesian coordinates,  with quantum numbers $n_x, n_y$,
or  polar coordinates  with  quantum numbers $n_\perp, m$.  We use the latter option.
Although  this is  standard  quantum mechanics, this set is less known. For completeness, we 
show how these functions are explicitly constructed 
and present several lowest functions explicitly for numerical use.

In cartesian coordinates, the 2-dimensional harmonic oscillator Hamiltonian and angular momentum along  the  $z$-axis read

 \bea
  \hat H&=& {1 \over 2 \mu} (\hat p_x^2+\hat p_y^2)+{\mu \omega^2 \over 2} (\hat x^2+\hat y^2)\nonumber\\
 \hat L_z&=&  \hat x \hat p_y- \hat y \hat p_x
\eea
     Since those operators commute, we will seek a common basis for  both using the ladder construction.
The operators reducing one quantum of oscillations are now defined as 

 \begin{equation}  a_R\equiv {1 \over \sqrt{2}}(a_x-i a_y),\,\,\,\,a_L\equiv {1 \over \sqrt{2}}(a_x+i a_y),
     \end{equation}
in terms of the one-dimensional operators $a_x,a_y$. These operators change the eigenvalue of $L_z$  or  $m$ by $\mp 1$ unit. 
Their hermitean conjugates change the eigenvalue by one more energy quantum.
Furthermore,
 
 \begin{equation}  
 \hat L_z/\hbar=  a_R^+ a_R - a^+_L a_L   
 \end{equation}
  and these two terms can be readily associated with the number of right or left rotating quanta.
  The energy is of course their sum, plus one from zero point oscillation
   \begin{equation}  E/\hbar \omega=  N_R+N_L+1   \end{equation}
From now on we will  use  the notation $n_\perp= N_R+N_L, m= N_R-N_L$
in reference to the  two quantum numbers of the states.
  
The explicit wave functions can be expressed in polar coordinates $\rho,\phi$
in which the reduction operators have the form

   \begin{equation}  a_R={1\over 2} e^{-i\phi} \big[\beta \rho + {1\over\beta} {\partial \over \partial \rho } 
   - {i \over \beta \rho} {\partial \over \partial \phi }\big]
    \end{equation}
       \begin{equation}  a_L={1\over 2} e^{i\phi} \big[\beta \rho +{1\over\beta} {\partial \over \partial \rho } 
   + {i \over \beta \rho} {\partial \over \partial \phi }\big]
    \end{equation}
where  $\beta=\sqrt{\mu\omega/\hbar}$ . Their hermitian conjugates are obvious.  Their actions  on the ground state 
   \begin{equation} \chi_{00}={\beta\over \sqrt{\pi}} exp\big(-\beta^2\rho^2/2\big)   \end{equation} 
 yield all excted states . For example, the state with maximal orbital momentum  has $m$ right rotating quanta $m=N_R,N_L=0$ 
    \begin{equation} \chi_{R^{NR}}={\beta\over \sqrt{\pi N_R!}} e^{i N_R \phi} (\beta \rho)^{N_R} exp\big(-\beta^2\rho^2/2\big)   \end{equation} 
    and the minimal one $m=-N_L$ follows using the  change $\phi \rightarrow -\phi, N_R  \rightarrow N_L $. 
    
    The  main advantage of this basis set is that the orbital momentum $\hat L_z$ commutes 
    not only with $H_0$ but with $V$ as well: so $m$ remains a good quantum number, before we consider the spin-flip residual interactions. Here are the next two $m=0$ functions (after $\chi_{00}$)
    \begin{eqnarray}  \chi_{RL}&=&{\beta\over \sqrt{\pi}} exp\big(-\beta^2\rho^2/2\big) (\beta^2\rho^2 - 1)\\
     \chi_{RRLL}&=&{\beta\over \sqrt{2\pi}} exp\big(-\beta^2\rho^2/2\big) (2-4\beta^2\rho^2 +\beta^4\rho^4)\nonumber\\
  \end{eqnarray}

  \subsection{The longitudinal harmonics}

Since the zeroth order Hamiltonian is  quadratic in $z^2 \sim \partial /
\partial p_z$, and $p_z\equiv x P$, the corresponding wave functions in $[0,1]$ are standing waves
 
 \begin{equation} f_{n_l}(x) \sim {\rm sin}(\pi (2 n_l-1) x ) \end{equation} 
that vanish at $x=0,1$ and symmetric in $x\leftrightarrow \bar x=1-x$. 
In addition, we also have the constant wavefunction $f_0(x)=1$, with the quantum number $n_l=0$.
Some matrix elements of this function are  logarithmically divergent at the endpoints, in which 
case pertinent but physical cutoffs will be needed.

\bibliography{mesons1X,string, allbib}

\end{document}